\documentclass[]{aa}  

\usepackage{graphicx}
\usepackage{txfonts}
\usepackage{amsmath}
\newcommand{\noop}[1]{}

\newcommand{\be}{\begin{eqnarray}}
\newcommand{\ee}{\end{eqnarray}}

\newcommand{\MSun} {\mbox{$M_{\odot}$}}

\begin{document} 

\titlerunning{Unbound stars hold the key to star cluster history}

\title{
%Dynamics adds challenges to membership determination\\ in young star clusters\\
Unbound stars hold the key to young star cluster history}

%\received{January 1, 2020}
%\revised{January 1, 2020}
%\accepted{\today}

   \author{Arunima Arunima \inst{1, }\inst{2}
          \and Susanne Pfalzner
          \inst{1,} \inst{2, } \inst{3}
          \and
          Amith Govind \inst{1, }\inst{2}
          }
             
\institute{J\"ulich Supercomputing Center, Forschungszentrum J\"ulich, 52428 J\"ulich, Germany\\
\email{s.pfalzner@fz-juelich.de}
\and Physics Department, University of Cologne, Cologne, Germany
\and Max Planck Institute for Radio Astronomy, Auf dem H\"ugel 69, 53121 Bonn, Germany
}

\date{Received ...}

\abstract
{}
{GAIA delivers the positions and velocities of stars at an unprecedented precision. Therefore, for star clusters, there exists much higher confidence in whether a specific star is a member of a particular cluster or not. However, membership determination is still especially challenging for young star clusters. At ages 2--10 Myr, the gas is expelled, ending the star formation process and leading to their expansion, while at the same time, many former members become unbound. As a first step, we aim to assess the accuracy of the methods commonly used to distinguish between bound and unbound cluster members; after identifying the most suitable technique for this task, we wish to understand which of the two populations is more suited to provide insights into the initial configuration and the dynamical history of a cluster starting from its currently observed properties.}
{Here, we perform N-body simulations of the dynamics of such young star clusters. We investigate how cluster dynamics and observational limitations affect the recovered information about the cluster from a theoretical perspective. }
{We find that the much-used method of distance and velocity cutoffs for membership determination often leads to false negatives and positives alike. Often observational studies focus on the stars remaining bound. However, bound stars quickly lose the memory of the pre-gas expulsion phase due to their ongoing interaction with their fellow cluster members. Our study shows that it is the unbound stars that hold the key to charting a cluster's dynamic history. Backtracking unbound stars can provide the original cluster size and determine the time of gas expulsion -- two parameters that are currently still poorly constrained. This information is lost in the bound population. In addition, former members are often better indicators for disc lifetimes or initial binary fractions. We apply the backtracking analysis, with varying success, to the clusters: Upper Scorpius and NGC 6530. For highly substructured clusters such as Upper Scorpius, backtracking to the individual subcluster centres will provide better results in future.}
{}

\keywords{stars: formation – open clusters and associations: general – ISM: clouds – solar neighbourhood}

\maketitle

\section{Introduction}
\label{sec:intro}

Star clusters are the nurseries for most stars \citep{Porras:2003,Lada:2003}. As such, young star clusters play a vital role in our understanding of how young stars form and develop. They signify the starting point for all that happens later on, as they provide the initial stellar mass distribution \citep[e.g.][]{Kroupa:2002} and the fraction of stars forming as a single-, binary-, or multiple-star system \citep[e.g.][]{Duchene:2018}.  
It is a standard procedure to use properties of clusters of different ages to obtain information on the dynamical development of young binary stars or the dispersal time of discs
\citep[e.g.][]{Haisch:2001,Ansdell:2017,Marks:2014,Ribas:2014,Richert:2018,Michel:2021}. 
Often the task of determining cluster membership and deriving the temporal development of specific properties are separate endeavours.
While distinguishing members is a challenge in itself, any bias in membership determination (i.e. false positives and false negatives) feeds through to the derived parameters used in other applications.

This study's central aim is to utilise cluster dynamics simulations to optimise the data used to determine a cluster's past.
Until recently, the role of dynamics during the formation history of young clusters was highly uncertain \citep[e.g.][]{Elmegreen:2000,Fujii:2012,Ward:2012,Banerjee:2017,Dib:2018}, mainly because observational limitations hampered precise velocity determination. 
The precision of data coming from the Gaia satellite \citep[][]{GAIA:2016,GAIA:2018,gaiaedr3_2021} helped shed light on this issue since a complete understanding of the dynamical evolution of present-day clusters has not been attained yet. 
Investigating a sample of 28 clusters and associations with ages \mbox{$\approx$ 1--5 Myr,} \citet{Kuhn:2019} found that at least 75\% of these systems are expanding at typical expansion velocities of the order of \mbox{$\approx$ 0.5 km s$^{-1}$.}
Cluster expansion was predicted by the gas expulsion scenario \citep[][]{Mathieu:1983, Lada:1984, Adams:2000, Kroupa:2001,Baumgardt:2007,Pelupessy:2012,Pfalzner:2013, Brinkmann:2017,Pfalzner:2021}.  
During the star formation phase, the stars are embedded in the gas and dust reservoir from which they are forming. %(Fig. \ref{fig:cluster_mass_radius}a). 
However, after approximately 1--2 Myr, the gas starts to be expelled from the clusters by various mechanisms
\citep[e.g. ][]{Krumholz:2009,Fujii:2021}.  Due to loss in gas and dust mass, the system is no longer in equilibrium. Therefore, a considerable portion of the stars, bound in the embedded phase, become unbound in the gas expulsion phase.% (\ref{fig:cluster_mass_radius}b).

 The three-dimensional information available from the \textit{Gaia} data has been a tremendous step forward in this field. Nevertheless, discriminating the members of star clusters  and associations from the foreground and background population is still challenging \citep[][]{Gagne:2018}. Many new methods have been developed for determining the members of open and globular clusters \citep[e.g.][]{Sollima:2019,Garro:2021,Vitral:2021MNRAS}. Cluster membership determination is challenging in the early expansion phase ($<$ 10 Myr), especially if a clear-cut distinction between currently bound and formerly bound (i.e. unbound) members is required.
In this case, there are additional difficulties to overcome compared to older clusters. 
First, the earliest stages of the formation of star clusters are hidden from view by gas and dust.  Thus, at this young age, veiling is a severe problem. 
Second, the young clusters' expansion requires additional attention in membership determination. Third, short- and long-lived clusters coexist during a 10 Myr timespan  \citep[][]{Lada:2003}. They undergo very different cluster dynamics \citep[][]{Pfalzner:2013}, and it is not always straightforward whether a specific cluster will remain bound for a long time or not. 

Here, we concentrate on these dynamical aspects of young short-lived clusters\footnote{The nomenclature of short-lived clusters is not unequivocal. While referred to as clusters while embedded, they are often classified as associations when the gas is expelled, and most of their stars become unbound. Here, we refer to short-lived clusters as clusters and point out expressly when talking about long-lived clusters,  that is, open and globular clusters.}.
 Any cluster observation is just a snapshot in time of the sequence of its dynamical evolution.
Based on simulations of the cluster dynamics, we show the importance of cluster dynamics in membership determination. We investigate the efficiency of backtracking cluster expansion and find that distinguishing between bound and unbound stars in the expansion phase is vital. Finally, we show that the unbound stars hold the key to determining a cluster's past.

\section{Cluster observation techniques}
\label{subsec:cluster_members}

Historically, star clusters have been identified visually as stellar density enhancements \citep{dreyer1888,trumpler1930,bailey1908,collinder1931}.
% ,mermilliod1995
Surveys like \textit{Hipparcos} \citep{hipparcos:1997},  2MASS \citep{2mass2006},  and \textit{Gaia}
%,cg2018,cg2019,cg2020,liupang2019,Sim2019,cantat_gaiadr2_2018,cantatgaudin_anders_2019
have each increased the samples by hundreds of candidate clusters. Due to \textit{Gaia}'s high-precision parallax measurements, the clustering of stars can be analysed in a higher dimensional space by combining their positions in the sky, proper motions, parallaxes, and radial velocities (when available). For studies which do automated blind searches with clustering algorithms, the youth of the stars is used as a confirmation of membership. 
%Observational studies of young clusters often start by identifying YSOs via 
Such youth indicators can be X-ray activity, infrared excess \citep{broos_etal2013,fiegelson_etal2013,getman_etal2017}, lithium abundance \citep{soderblom2010}, and gravity-sensitive spectral indices such as TiO molecular lines \citep{wilking2005}, empirically constructed spectral indices \citep{damiani_etal2014}, or the shape of the $H$-band peak \citep{scholz_etal2009}.

%Recently, a number of approaches have been used for automating cluster searches and assigning membership. 
%Some studies use only photometry \citep{Buckner:2013}, only astrometry \citep{Sampedro2017,cantat_gaiadr2_2018,Crundall2019,pang2020,Pang2021,chronostar2} or combination of both photometry and astrometry \citep{kronemartins_moitinho2014,wright_etal2019,Wright:2019,cg2020} data. 
Among the clustering algorithms, one can distinguish different classes: Density-based spatial clustering like DBSCAN  \citep{Ester96DBSCAN,Wilkinson2018,Zari2019,cg2019,cg2020,cg2021,Hunt2021}, HDBSCAN \citep{campello2013hdbscan}, and OPTICS (Ordering Points To Identify the Clustering Structure; \citealt{Ankerst99OPTICS}),  multidimensional Gaussian-based methods \citep{Vasiliev2019,cantatguadin2019perseus,kuhn_etal2020}, $k$-means clustering \citep{macqueen1967kmeans,Hunt2021}, and Friend of Friend algorithm (FoF; \citealt{liupang2019}). In addition, there exist several unsupervised algorithms like UPMASK  \citep{kronemartins_moitinho2014,cantatgaudin_etal_2018,cantatgaudin_anders_2019}, the nearest neighbour-based method by \citet{heNN2021}, and STARGO \citep{tang2019,zhang2020,pang2020}.

Young star clusters pose additional challenges compared to open or globular clusters due to their highly dynamic nature after gas expulsion. Although space velocity is used to identify clusters, algorithms rarely consider dynamics.  Observations only provide a snapshot in the dynamic evolution of the cluster. Hence, even clustering in the velocity space at the present moment might be a chance alignment as the velocity changes rapidly in young star cluster members. 
More limitations in identifying clusters come from \textit{Gaia}'s poor completeness in crowded fields and no particular regard for binarity. Moreover, young clusters are still embedded in natal gas and dust that can not be penetrated by optical wavelengths, which presents another difficulty in identifying and analysing young clusters. 

\citet{blaauw1964} first gave the notion of linear expansion in associations, assuming that all members move away from their birthplace without any forces acting on them. 
Then, the reciprocal of the expansion coefficient can provide an estimate of the association's kinematic age. Alternatively, the individual motions of the stars can be traced back until they reach the smallest configuration at a past time, and the kinematic age, as well as the initial configuration of the association, can be possibly obtained \citep{blaauw1978}.

Most studies apply cutoffs to remove objects with low-quality astrometry and outliers. 
The sigma-clipping method aims to reduce the chances of contaminants or uninformative stars and improve clusters' signal-to-noise ratio (S/N). Alternatively,  outliers can be modelled in the fitting procedure without rejecting points a priori (see \citealt{hogg2010}).

%This simplified theory was considered further and the effect of the Galactic potential on the orbits of the stars was included to be an epicyclic approximation. However, \citet{brownetal1997} combined $N$-body simulations with the effects of the Galactic field from the epicyclic approximation and concluded that for the typical velocity dispersion in associations, the effects of the Galactic tidal field do not substantially contribute to the error in kinematic age determination.  

%With the availability of \textit{Hipparcos} data, efforts were made to study the kinematic evolution of the nearby associations . However, 

Before \textit{Gaia}, the significant errors in astrometry and the low number of confirmed members with available radial velocities were the main hindrances in the analysis \citep{fernandez2008}. 
The higher precision of the \textit{Gaia} data allows for better trace-back analysis. For example, recent studies by \citet{Heyl:2021a,Heyl:2021b} trace back the stars of clusters aged 40--200 Myr using \textit{Gaia} EDR3 data and determine their kinematic ages. Similarly, \citet{Schoettler2021} trace back  runaway (RW) and slower walkaway (WW) stars within a distance of 100 pc of NGC 2264 to the three subclusters S Mon, IRS 1 and IRS 2. The study by \citet{Ma2022} uses \textit{Gaia} DR2 data to trace back (and extrapolate) the trajectories of members of the Scorpius-Centaurus (Sco-Cen) association and find evidence of past and future close stellar flybys. 
%Having complete 6-dimensional astrometric data makes the trace-back more precise and recent developments like \texttt{Chronostar} in \citet{Crundall2019} are improving the methodology further.

%Young clusters are highly dynamic and any cluster observation is just a snapshot in the time sequence of strong dynamic evolution of the cluster. 

Observational challenges like distinguishing the cluster population from the back and foreground stars, limiting magnitudes, imprecision of derived properties like age and mass, etc., complicate backtracking. Here we apply backtracking to snapshots in the simulations of the cluster dynamics. Under these idealised conditions, membership is certain, the exact positions and velocities of the stars are known at all times, and last, but not least, we know what the result should be. 
This certainty allows us to determine the most expedient method and suggest measures to optimise the backtracking technique.

\section{Cluster simulation method}

We use a sub-set of simulations of the dynamics of clusters containing $N$ stars we performed recently \citep{Pfalzner:2021}, using the simulation code \texttt{NBODY6++GPU} \citep{Aarseth:2003}. The simulations try to represent the situation in real clusters as closely as possible by adopting initial conditions backed by recent observations and following the observed cluster expansion derived from the sizes of clusters in the age range of 1--10 Myr. Here we give only a summary of the assumptions, and the numerical method we applied in \citet{Pfalzner:2021}, as the actual choice of simulation parameters is uncritical for the general challenges in membership determination and backtracking of the cluster history.

We model the dynamics of the young clusters covering all the phases:
Starting from the embedded phase, we simulate the subsequent gas expulsion that leaves the cluster in a super-virial state and results in the cluster expanding until it reaches a new equilibrium. It is assumed that all stars are already formed and that the gas expulsion occurs at $ t$\textsubscript{emb} = 2 Myr. Observations indicate that the entire gas expulsion process takes $\approx$ 1 -- 2 Myr \citep[][]{Kuhn:2019}. Simulations investigating the dependence of the cluster dynamics on the gas expulsion time found that the gas expulsion can be modelled as being instantaneous  \citep{Geyer:2001,Portegies:2010}. Stellar evolution has not been included in this work as it has little influence on the results. 

We analyse the dynamics of clusters with different numbers of cluster members $N$. The corresponding clusters' masses $M$\textsubscript{c} and sizes, illustrated by their half-mass radius $r$\textsubscript{hm}, are given in  \mbox{Table \ref{tab:cluster_parameters}.} Low-mass clusters are usually smaller than high-mass clusters of the same age \citep{Lada:2003,Adams:2010,Pfalzner:2016}. This relation between the cluster's mass and its half-mass radius can be approximated by a power law:
\begin{equation}
\label{eq:mass-radius relation}
    M_c = C r\textsubscript{hm}^\gamma.
\end{equation}
The values of the constant $C$ and scaling exponent $\gamma$ differ in different observational studies due to the involved observational uncertainties. The clusters' sizes given in Table \ref{tab:cluster_parameters} are based on the mass-radius relation by \cite{Pfalzner:2016} where $C = 717.794$ and $\gamma = 1.7 \pm 0.2$.
% The choice of cluster size is based on the corresponding observational data. 
We assume that the star formation efficiency in the system is 30 \%  \citep{Lada:2003}, which sets the gas mass. The gas and dust component of the embedded phase is implemented as a background potential. 

% Although this value has an observational basis (Lada & Lada, 2003), the SFE values in individual clusters might vary between 10\% to 35\%. Even within a cluster, the SFE might depend on the local stellar density and hence, the local values can be higher in the central parts (Moeckel & Bate, 2010; see also: Parmentier & Pfalzner, 2013).
%
\begin{table}[b!]
\caption{Initial cluster parameters for the simulation campaign using mass-radius dependencies.}
%        \centering
\begin{tabular}{rrrlrc}
        \hline
        \hline
%        \begin{tabular}[c]{@{}c@{}}Model\\ {} \end{tabular} &
        \begin{tabular}[c]{@{}c@{}}N\\ {} \end{tabular} & \begin{tabular}[c]{@{}c@{}}N\textsubscript{sim}\\ {} \end{tabular} &
        \begin{tabular}[c]{@{}c@{}}M\textsubscript{c}\\ {[}M$_{\odot}${]}\end{tabular} & \begin{tabular}[c]{@{}c@{}}r\textsubscript{hm}\\ {[}pc{]}\end{tabular} & \begin{tabular}[c]{@{}c@{}}M\textsubscript{t}\\ {[}M$_{\odot}${]}\end{tabular} &  \begin{tabular}[c]{@{}c@{}}t\textsubscript{emb}\\ {[}Myr{]}\end{tabular}\\
        \hline
        200 &           1941 &     117.99 &  0.26 &    393.31 & 2.0 \\
        1000 &           497 &     589.97 &  0.67 &   1966.57 & 2.0 \\
        4000 &      127 &  2359.88  & 1.3 &  7866.27 & 2.0 \\
        % 10000&            95 &    5899.71 &  1.3  &  19665.68 & 2.0 \\ 
                   \hline
               \end{tabular}
               	\hspace{0.5em}
	\tablefoot{Here  $N$ denotes the number of cluster members,  $ N$\textsubscript{sim} the number of simulations, $ t$\textsubscript{emb} the duration of the embedded phase, $M$\textsubscript{c} the stellar mass of the cluster, $r$\textsubscript{hm} the half-mass radius, and $M$\textsubscript{t} the total cluster mass  (stars + gas).}
        \label{tab:cluster_parameters}
\end{table}

In our simulations, a test particle represents a star with a given mass, position, and velocity. The particles' positions are chosen so that the resulting stellar number density distribution obeys a King profile with King parameter, $W_0 = 9 $ \citep{King:1966}.  The King model is an empirical law that can not be defined analytically. It consists of an energy distribution function of the form
\begin{equation}
    f_K (\mathcal{E}) = \begin{cases} 
      \rho_1(2\pi\sigma^2_K)^{-3/2}(e^{\mathcal{E}/\sigma^2_K}-1) &  : \mathcal{E} > 0,\\
      0 & :  \mathcal{E} \leq 0,
   \end{cases}
\end{equation}
with $\mathcal{E} = \Psi - \frac{1}{2}\nu^2$ and $\Psi = -\Phi + \Phi_0$ being the relative energy and relative potential of a particle, respectively. Also, $f(\mathcal{E}) > 0$ for $\mathcal{E}>0$ and $\sigma_K$ is the King velocity dispersion. The profiles are characterised by the King parameter $W_0 = \Psi/\sigma^2_K $, an increase of which signifies decrease in the relative size of the cluster core $r_c/r\textsubscript{hm}$. Observationally, determining the stellar density distribution of young star clusters can be challenging but it has been found that young clusters are best represented by King model with $W_0 \geq 7$ \citep{hillenbrand1998,nurnberger2002}. The choice of $W_0$ mainly affects the size of the central high-density area. Hence, the number of expelled stars also depends on the choice of $W_0$. Even for a relatively steep $W_0$ = 9-potential, the number of escapers is $<$ 1\%. Therefore,  the conclusions about membership determination methodology are unaffected by the choice of potential. 
The individual test particles are assigned masses following the initial mass function (IMF) by \citet{Kroupa:2002}, with the lower mass limit set to 0.08 $M_{\odot}$ (hydrogen-burning limit) and an upper mass limit of 150 $M_{\odot}$. 
Potentially existing initial mass segregation in the clusters is neglected.
The cluster members are given velocities following a Maxwellian distribution. We assume that the cluster is initially in virial equilibrium. 

We perform ($N\textsubscript{sim}$) simulations for every cluster mass, where the actual distribution of the stars depends on the seed selected in the randomised procedure. We analyse all the simulation results in this statistical study.  However, why a specific method works or fails,  we illustrate exemplarily for just one specific randomly chosen realisation in Figs. 1 -- 3. Figures 6 -- 8 also show the method applied to randomly chosen specific clusters for visual understanding; however, statistical results are mentioned in the text.

For simplicity, we exclude primordial binaries, modelling all cluster stars as initially being single stars. The absence of primordial binaries can lead to underestimating ejections from the cluster centre \citep{Heggie:1975}. However, in most clusters, $\ll$1\% of the stars are affected \citep{Olczak:2006}.

%
%Each method has its own advantages and disadvantages in identifying clusters or subclusters and rejecting non-clustered field stars depending on the various assumptions used. This introduces biases in the membership and hence, derived properties of these resulting groups. Some of the main methods used in literature are discussed below along with the possible biases they might introduce.

\begin{figure}[t]
\includegraphics[width=0.5\textwidth]{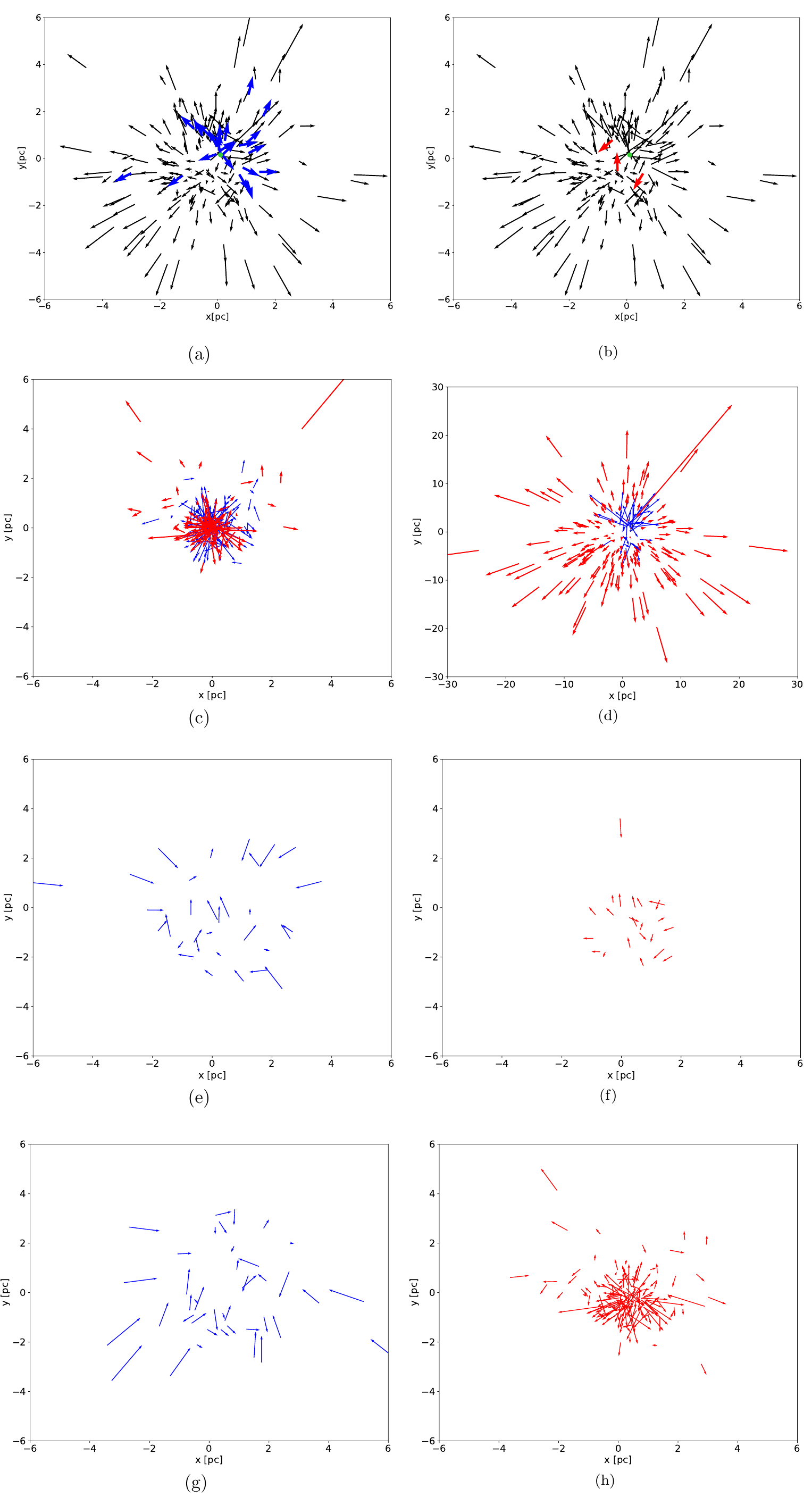}
\caption{Snapshot of the positions and velocities of example simulations with $N$ = 200. Velocity vectors of bound stars are highlighted in blue, and those of unbound stars in red. Counter-intuitive examples of (a) outward-pointing distant bound stars and (b) inward-pointing central unbound stars. Snapshot of the temporal development at (c) t=2 Myr and (d) t=10 Myr. Backtracking from the results at 10 Myr to 2 Myr considering only the stars within 6 pc from the cluster centre for (e) bound stars only and (f) unbound stars only. Same backtracking considering all the (g) bound stars and (f) unbound stars of the cluster. A film of the cluster dynamics and the backtracking can be found at https://doi.org/10.5281/zenodo.6041920
%Green dot shows the centre of mass of the cluster.
}
\label{fig:Velocity_vectors}
\end{figure}

\begin{figure*}[t]
    \centering
    \includegraphics[width=0.98\textwidth]{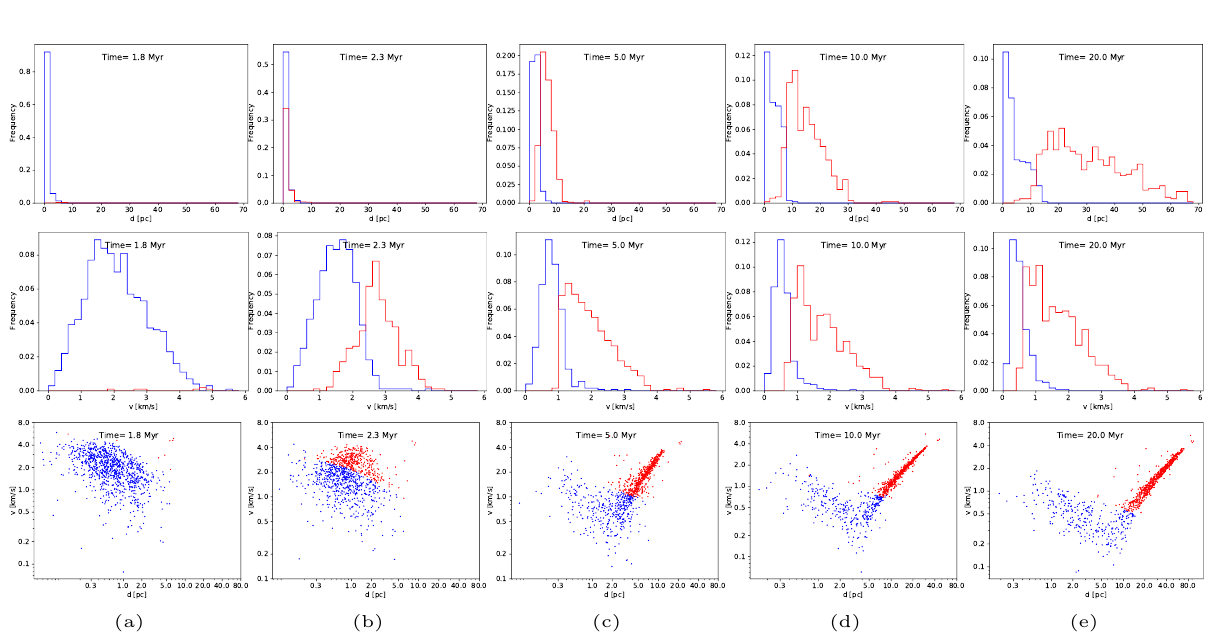}
    \caption{Snapshot of distance (top) and velocity distribution (middle), and distance vs velocity scatter plot (bottom) (a) before gas expulsion ($t=$ 1.8 Myr), (b) just after gas expulsion ($t=$ 2.3 Myr), (c) at $t=$ 5 Myr, (d) at $t=$ 10 Myr, and (e) at the end of our simulation ($t=$20 Myr). All plots show the bound stars in blue and the unbound stars in red. A simulation of $N=1000$ stars is used here.}.
    \label{fig:distributions}
\end{figure*}

\section{Results}

Observations investigate one specific cluster at a snapshot of its development. Mimicking this observational situation,  we randomly choose one of our sets of simulations and investigate it at a specific time. However, unlike actual observations, we have complete temporal information available. Hence, we know the past and the future of this particular cluster down to the path of each star. Equally, all other observational challenges, like membership uncertainty due to back and foreground populations and limiting magnitudes, are removed. We even know each star's exact properties like its mass, position, and velocity. This information allows us to investigate the fundamental and unavoidable challenges in backtracking caused by the cluster dynamics that exist even without the mentioned additional observational difficulties.

%\begin{figure}[h]
%    \includegraphics[width=0.5\textwidth]{backtracing.pdf}
%\caption{Snapshot of an example of our set of simulations a) at t=2 Myr and b) at 10 Myr. For each star the velocity vector is shown, where the bound stars are indicated in blue and the unbound stars in red. Backtracking from the results at 10 Myr to 2 Myr considering only the stars within 6 pc from the cluster center for (c) bound stars only and (d) unbound stars only. The model cluster contained $N$ = 200 stars. A film of the cluster dynamics and the backtracking can be found at \textcolor{red}{eventually add link.}[DOI:10.5281/zenodo.6041920 (not published yet)]}
%\label{fig:back_tracing}
%\end{figure}

\subsection{Bound and unbound stars}

After gas expulsion, bound and unbound stars coexist in the same spatial area for some time. Distinguishing the two populations is vital for some applications; it does not matter or is not even desirable for others. An example of the latter is the use of clusters in determining disc lifetimes \citep[][]{Haisch:2001}. Here, it is best to identify all stars that once formed together in the cluster. However, if one is interested in the long-term development of clusters ($\gg$ 20 Myr), one would be predominantly interested in the portion of stars that remain bound. We subsequently see here that using backtracking to distinguish between bound and unbound stars after gas expulsion is the key to success in obtaining valuable information concerning a cluster's past. At each snapshot of the simulations, bound and unbound stars are defined as those having positive and negative total energy respectively. However, in observations, distinguishing between these two states is often not straightforward. 

\subsubsection{Velocity vectors}

Individual stars are sometimes classified as bound or unbound simply because their velocity vectors point towards or away from the cluster centre.  In the past, doubts about this approach were usually anchored on the fact that only two-dimensional information was available. However, even with three-dimensional information becoming more accurate, this method is not advisable even for perfectly known 3D velocities for the following reason: The top row of Fig. \ref{fig:Velocity_vectors} shows a typical snapshot of a randomly chosen example from our sample of simulated clusters. The cluster centre is marked as a green dot as a reference point. As the many outward-pointing velocity vectors indicate, this cluster is in the expansion phase, with many former members becoming unbound. Nevertheless, a considerable fraction of the outward-pointing velocity vectors belongs to stars that remain bound in the long term. Examples of such stars are shown in blue. Equally, stars that point inwards and are close to the cluster centre can nevertheless be unbound (shown in red). The dynamics of these example stars can be seen better in the corresponding video at \url{https://doi.org/10.5281/zenodo.6041920}.   Especially among the bound stars with outward-pointing velocity vectors, quite a few are bound despite being located at relatively large distances from the cluster centre. We find that  there is a high failure rate in this approach, not only for this specific cluster, but for all clusters in our extensive sample. The situation improves for clusters aged more than 15 Myr as many of the unbound stars are better identifiable by their larger distances to the cluster centre.

\begin{figure}[h]
\includegraphics[width=0.48\textwidth]{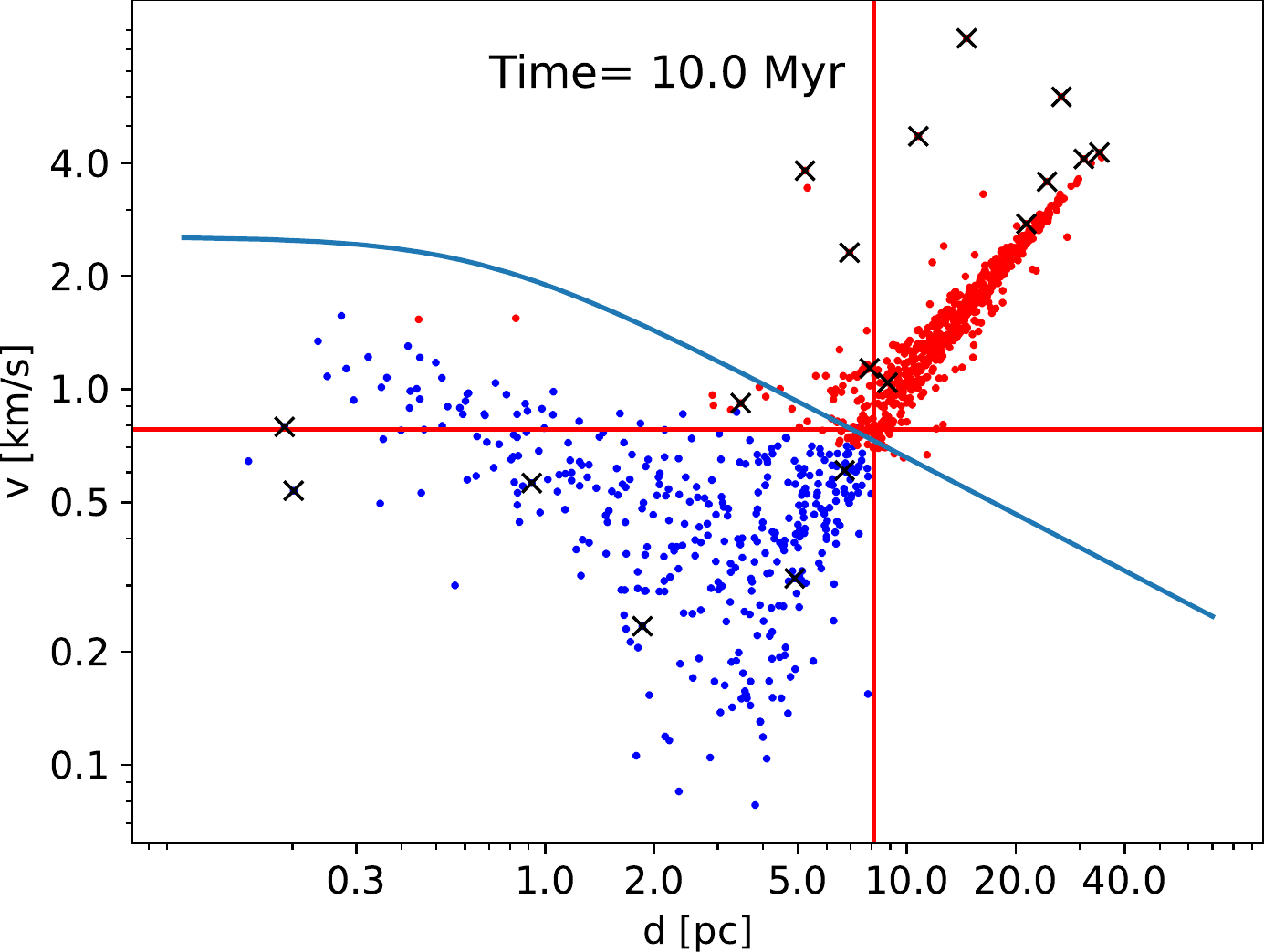}
\caption{Phase space diagram for an $N=1000$ star cluster simulation at $t=10$ Myr. The bound and unbound members are shown in blue and red colours respectively. Vertical and horizontal red lines indicate distance and velocity cutoffs respectively for unbound stars. The light blue line represents the analytical escape velocity dependence on distance from the cluster centre derived assuming a Plummer distribution for the members. The black crosses show the stars that underwent a strong encounter.}
\label{fig:cut-off}
\end{figure}

\subsubsection{Advantage of using unbound stars for backtracking}

The size of a cluster before expansion sets in is an essential parameter for constraining the cluster formation process. Besides the density profile, the size of the cluster core and half-mass radius are good indicators of the cluster density and, thus, the importance of the environment in the star and planet formation process. The environment's influence includes close stellar flybys and external photo-evaporation that can truncate protoplanetary discs or completely destroy them \citep{Vincke:2015, Winter:2018,Concha:2019}. These processes influence the type and frequency of the formed planetary systems. Another example is binary capture and destruction processes which can alter the binary fraction in clusters \citep[][]{Kaczmarek:2011, Marks:2014,Guszejnov:2022}.

We find that using just the unbound stars gives the best result in determining the pre-expansion cluster size. As an example, the second row in Fig. \ref{fig:Velocity_vectors} illustrates the cluster expansion by showing the bound and unbound stars, including their velocity vectors,  (a) shortly after gas expulsion and (b) at 10 Myr for a cluster with $N$ = 200. We note the different scales. Using only the bound stars for backtracking (see Fig. \ref{fig:Velocity_vectors}g) results in a relatively poor constraint on the pre-expansion size. The best performance is obtained using only the unbound stars (see Fig. \ref{fig:Velocity_vectors}f). The reason is twofold: First, the velocity vectors of the unbound stars are rarely altered after gas expulsion. By contrast, bound stars quickly lose the memory of the pre-gas expulsion phase due to their ongoing interaction with their fellow cluster members. In particular,  close encounters hinder efficient backtracking for the bound stars. Second, there is a more significant number of unbound than bound stars. Thus, statistical uncertainties are more easily averaged out.

\begin{figure}[t]
    % \centering
    \includegraphics[width=0.48\textwidth]{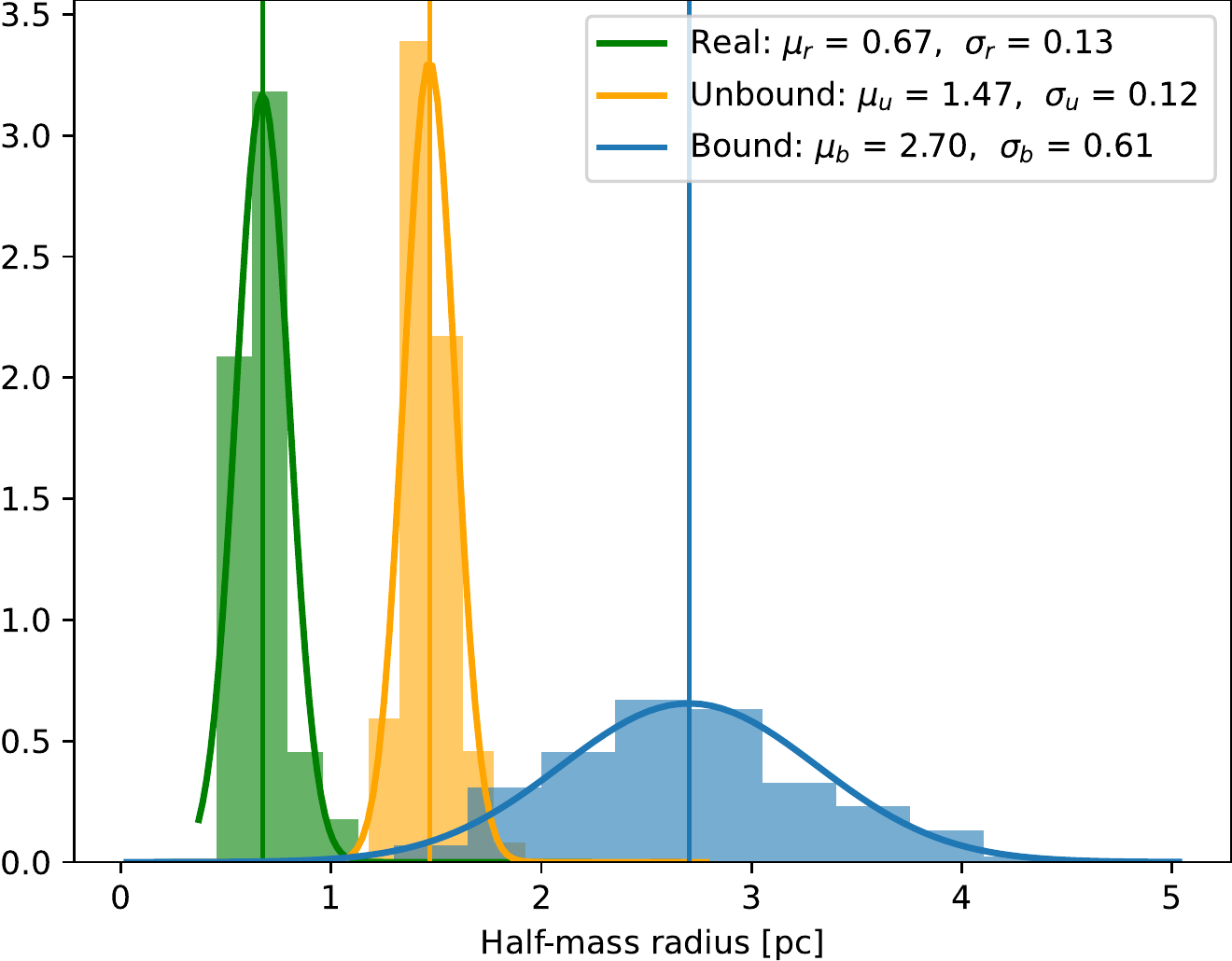}
    \includegraphics[width=0.48\textwidth]{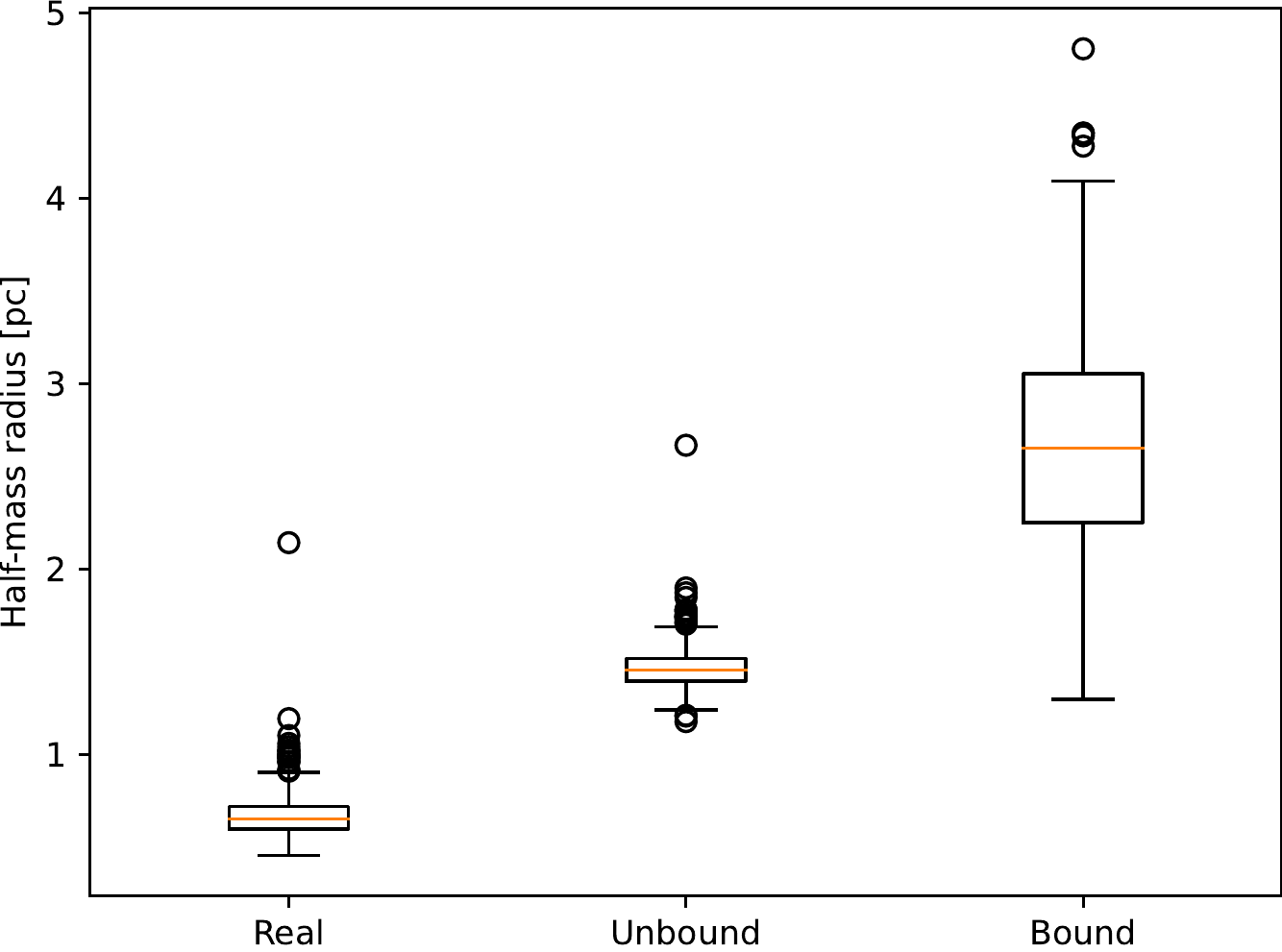}
    \caption{Distributions of sizes derived using actual positions of all stars (Real, shown in green), using backtraced positions of unbound stars (Unbound, shown in orange), and using backtraced positions of bound stars (Bound, shown in blue) shown with histograms (top) and boxplots (bottom). The box extends from the lower to upper quartile values of the data, with a line at the median while the whiskers reach 1.5 times the interquartile range from the box.}
    \label{fig:size}
\end{figure}

Figure \ref{fig:size} gives a more quantitative idea of the use of bound vs unbound stars for backtracking and deriving the pre-expansion cluster size. All the simulations of $N=1000$ cluster have been used to obtain these distributions. It can be seen that the size distribution obtained using unbound stars is closer to the real size distribution than the size distribution obtained using bound stars. Performing a t-test on the two size distributions with the null hypothesis being that the distributions have the same mean---while the alternative hypothesis is that bound stars have a larger mean than unbound stars---results in a $p$-value much lower than the significance level $\alpha=0.01$.
% Considering the sizes derived using bound and unbound stars to be samples from different populations (which is justified by very low p-value in KS test), the probability of a random variable drawn from the bound size population turning out to be greater than a random variable drawn from the unbound size population is found to be 97.6\%. Not only is the mean of the sizes obtained using unbound stars closer to the mean of the real sizes, their standard deviation is much less than that of the sizes obtained using bound stars. 
Hence, unbound stars are clearly better at recovering the size of the cluster before gas expulsion than bound stars.

\subsubsection{Distance and velocity cutoffs for bound-unbound classification}

While distinguishing between the bound and unbound population is straightforward in simulations,  it is very challenging in observations. Often a cut in the distance to the cluster centre or the velocity is used to distinguish between bound and unbound stars. Here we want to test when such a method is successful. 

In our simulation, the relevant time frame starts at 2 Myr, when the gas expulsion happens, and many stars become unbound. Figure \ref{fig:distributions} shows snapshots of the distributions of the stellar distance to the cluster centre and velocity distribution before (1.8 Myr), just after gas expulsion at 2.3 Myr, during the expansion process (5 and 10 Myr) and towards the end (20 Myr) of the expansion phase for an example cluster. The distributions for the bound (blue) and unbound (red) stars are shown separately. As we chose the cluster to be in virial equilibrium, very few stars become unbound before gas expulsion (see Fig. \ref{fig:distributions}a). The few unbound stars during this phase result from close encounters leading to ejections. 
However, after gas expulsion, many stars become unbound. Bound and unbound stars share considerable parts of the phase space for quite some time, as seen in the bottom row of Fig. \ref{fig:distributions}. This increases the complexity of making the distinction. 

\begin{figure}[h]
\includegraphics[width=0.48\textwidth]{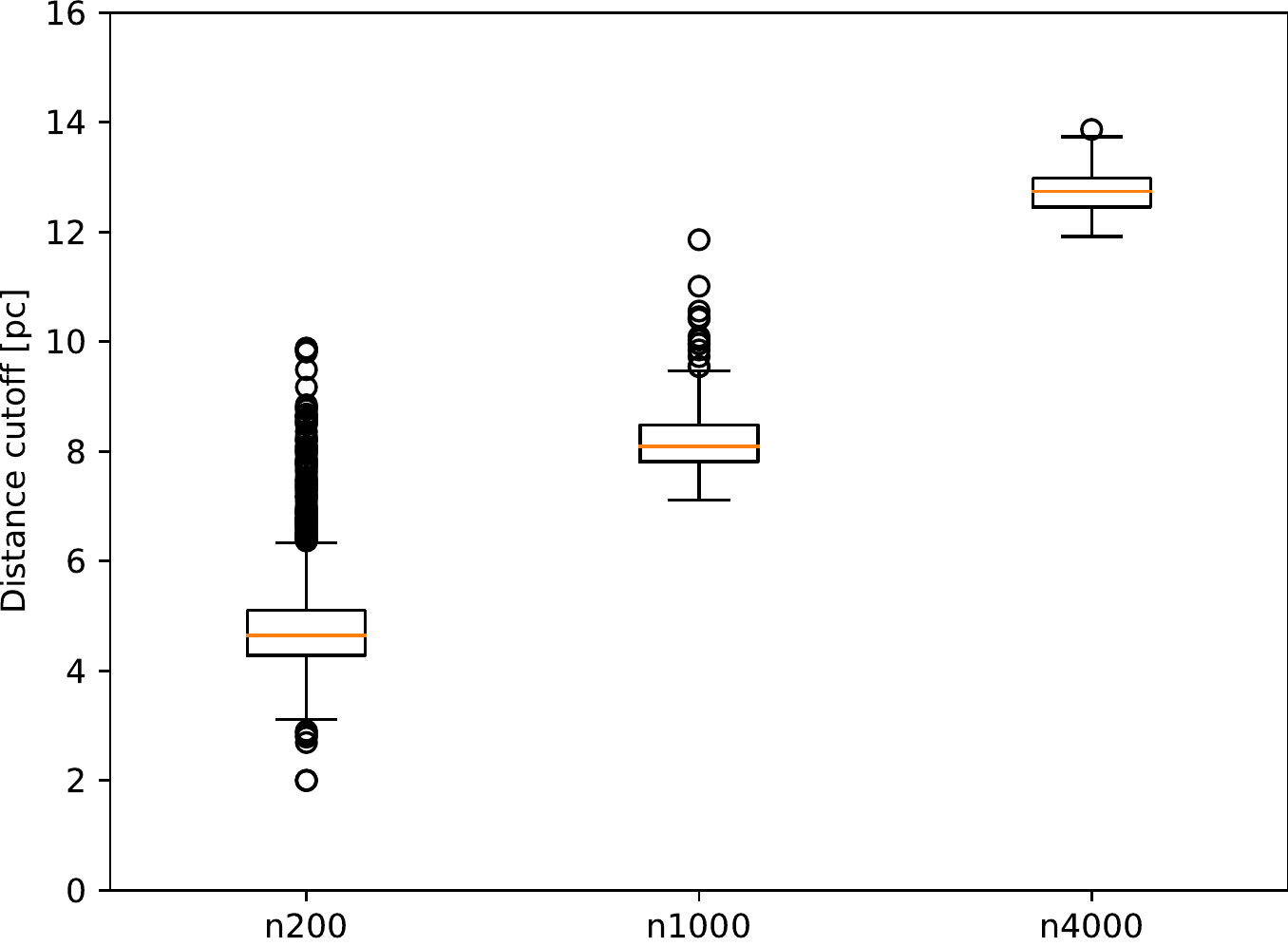}
\includegraphics[width=0.48\textwidth]{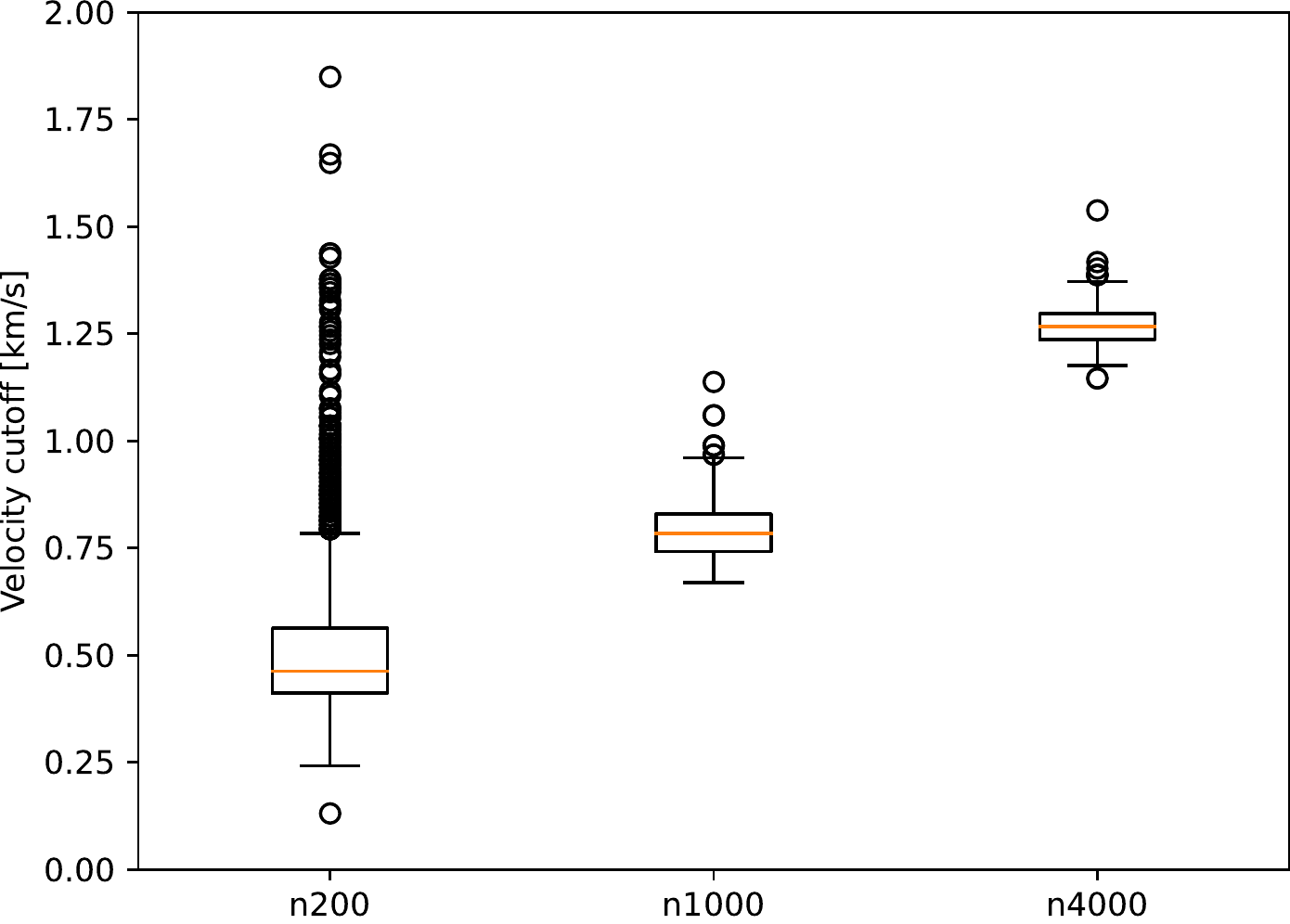}
\caption{Distance (top) and velocity (bottom) cutoffs for selection of unbound members for clusters with different number of members: $N=200$, 1000, 4000. The box extends from the lower to upper quartile values of the data, with a line at the median while the whiskers reach 1.5 times the interquartile range from the box.}
\label{fig:cutoff_boxplot}
\end{figure}

In observations, usually, a velocity cutoff is chosen as a given deviation from the mean for making this distinction \citep[e.g. ][]{luhman2018tau,Bastian2019,Esplin2019}. However, the location of these cutoffs is not apparent. Thus, there is some element of arbitrariness here, and this is even more so for distance cutoffs. However, in our simulations, we are in the ideal situation where we can determine where to apply the cutoff in distance and velocity. These experiences can be used to provide guidelines for both types of cutoffs. Figure \ref{fig:cutoff_boxplot} shows suggestions for the choice of distance and velocity cutoff for clusters older than 5 Myr. These have been calculated to minimise the sum of the false positive rate (FPR) and false negative rate (FNR) for all the simulations. 

It does not make much sense to make distance and velocity cutoffs in clusters younger than at least \mbox{5 Myr} to avoid substantial errors in the classification of the members. However, even at 5 Myr, the FPR and FNR introduced by a cutoff can be of the order of 15\% -- 30\%. Generally, the percentage of stars identified as bound members while being unbound is higher than the opposite situation. Only for clusters older than 10 Myr, this method is relatively robust as the overlap in phase space is of the order of 5\% -- 10\%. Figure \ref{fig:cut-off} shows the phase space diagram for a simulated cluster of 1000 stars with red lines at a distance of 8.09 pc and a velocity of 0.78 km/s representing the distance and velocity cutoffs shown in Fig. \ref{fig:cutoff_boxplot}. Applying these to the distribution of all simulations of 1000 stars leads to a median FNR of 9.7\%. The 25th and 75th percentile of the distribution of FNR are 7.5\% and 11.4\%, respectively. We represent this as an FNR of $9.7_{-2.2}^{+1.7}\%$. Similarly, an FPR of $0\pm0\%$ is obtained. The percentage of correctly identified stars is found to be $94.1\pm1.1$\%.

Combining distance and velocity cutoffs gives the best distinction. This can be done by analytically determining the dependence of the escape velocity of the stars on the distance from the cluster's centre. Although the distribution of the stars in the simulations follows a \cite{king1966} profile, we use an approximation of a \cite{plummer1911} profile to obtain an analytical solution. The escape velocity  $v$\textsubscript{esc}$(r)$ at any point in the cluster is then described by 

\be
    \label{eq:v_escplummer}
    v\textsubscript{esc}(r) = \sqrt{\frac{2GM\textsubscript{cl}}{\sqrt{a^2 + r^2}}},
\ee

where $M$\textsubscript{cl} is the cluster mass, and $a$ is the initial half-mass radius.
This analytical cutoff can be seen in Fig. \ref{fig:cut-off} as the blue curve. Applying this as the cutoff for bound-unbound star distinction leads to an FPR of $0.74_{-0.47}^{+0.91}$\% and an FNR of $4.80_{-0.10}^{+0.88}$\%. The median of the distribution of the correctly identified stars' percentage is found to be $96.7_{-0.7}^{+0.5}$\%. Hence, this analytical cutoff is an improvement over the distance and velocity cutoffs in the case of our simulations.

\begin{figure}[t]
    % \centering
    \includegraphics[width=0.5\textwidth]{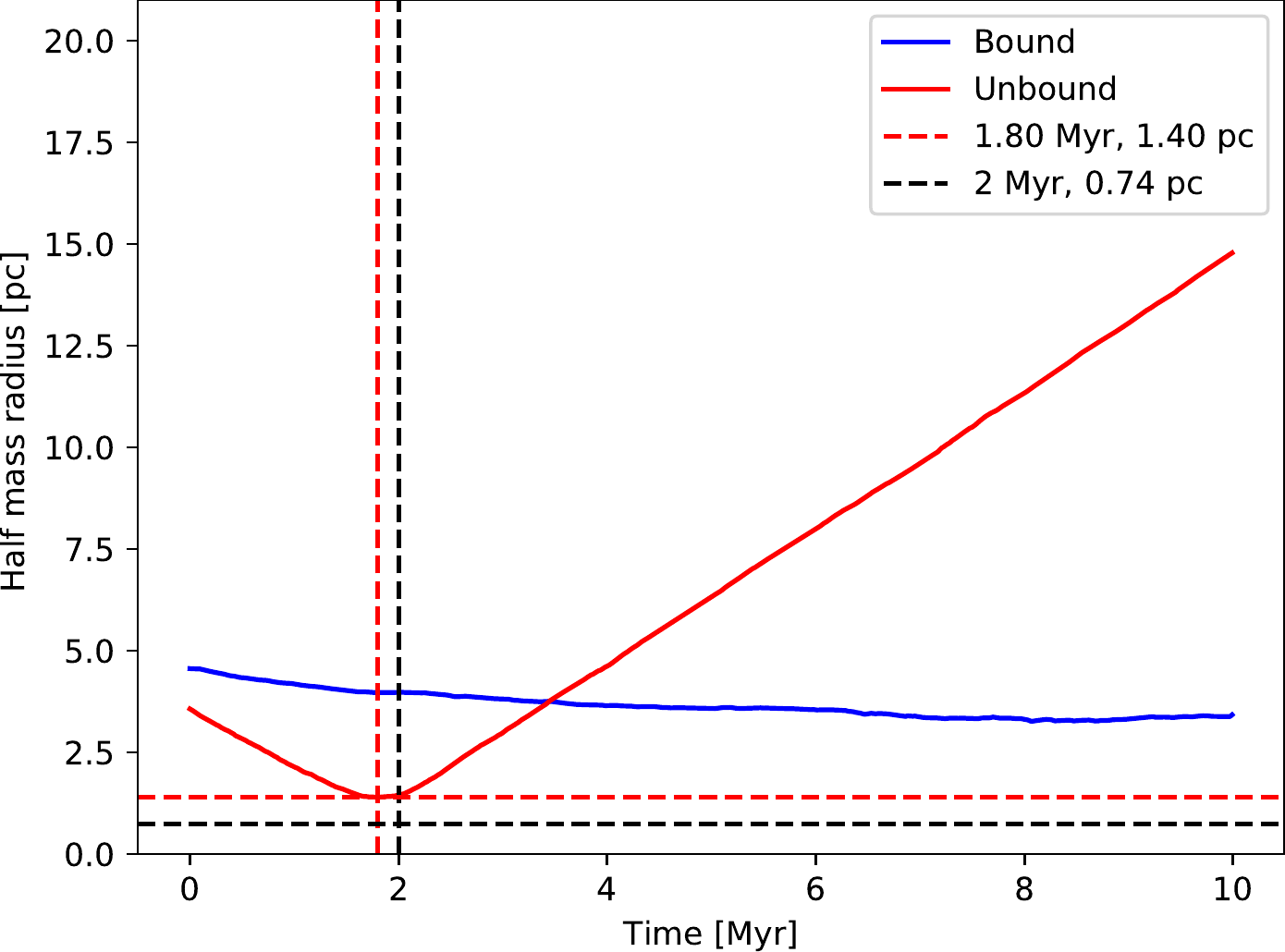}
    \includegraphics[width=0.5\textwidth]{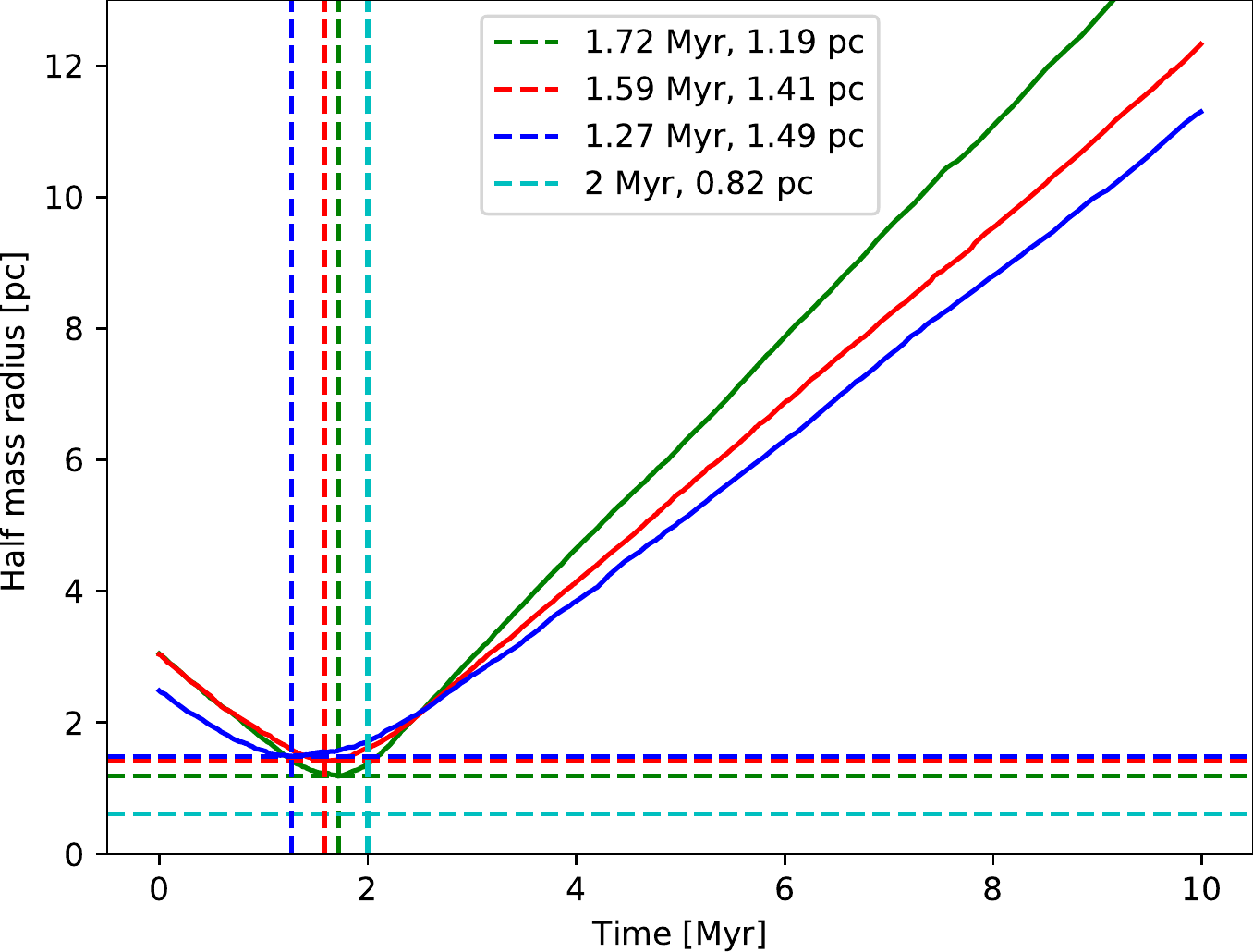}
    \caption{Backtracked half-mass radii for a simulation with 1000 stars, Top: using bound (blue) and unbound (red) members only. Red dashed lines show $t$\textsubscript{emb} and $r$\textsubscript{hm} at the time of gas expulsion determined using unbound stars whereas black dashed lines show the actual values of the same. Bottom: using unbound stars within 10 pc (blue), 20 pc (red) and 40 pc (green) from the cluster centre. The actual values of $t$\textsubscript{emb} and $r$\textsubscript{hm} at the time of gas expulsion are shown in cyan.}
    \label{fig:cut-off-trace-back}
\end{figure}

\subsection{Backtracking}

In the following, we use our simulations of the cluster dynamics to develop guidelines for backtracking depending on cluster type, age, and mass. We subsequently demonstrate that using the right subset of stars for backtracking is the key to making the most of the available information. Here, we employ the simplest form of backtracking, namely, taking present-day positions and velocities as constant values and just reversing the arrow of time (i.e. neglecting any source of acceleration acting upon the stars).

The high quality of the recent \textit{Gaia} data allows backtracking from the observed present situation holding the promise to reveal information about a cluster's past. So far, unbound stars are chiefly analysed as `runaway' (\mbox{$v >$ 30 km/s}) stars and `walkaway' (\mbox{5 km/s $< v <$ 30 km/s}) stars \citep{Eldrige:2011,Schoettler:2020}. The idea is that both types of high-velocity stars have been ejected from their star-forming regions, and backtracking will allow us to determine their origins and characterise their parent star cluster  (e.g. \citealt{Olczak2008,Farias:2020,Schoettler2021}). 
% Based on \textit{Gaia} EDR3, recent examples, backtracking in four other open clusters with ages between 40--200 Myr \citep[][]{Heyl:2021b} to determine the kinematic age of these clusters. 
\citet[][]{Schoettler2021} search for runaway and walkaway stars within 100 pc of the 3--5 Myr old cluster NGC 2264 using \textit{Gaia} DR2. They compare the number of the runaway and walkaway stars (17) to a range of N-body simulations with different initial conditions and find consistency with initial conditions with a high initial stellar density \mbox{($\approx$ 10 000 \MSun\  pc$^{-3}$)} and a high initial amount of spatial substructure.

However, our simulations find that high-velocity ejections are rare for short-lived clusters. We found no ejections with\mbox{ $v >$ 30 km/s} and only a few with  \mbox{$v >$ 5 km/s}. Thus, backtracking based on runaway and walkaway stars suffers from low-number statistics for young clusters ($<$ 20 Myr) typical for the solar neighbourhood. As the ejection happens mainly from the highest-density regions of the cluster, the derived age at gas expulsion is too short, and the cluster size is also too small. For the much denser clusters that turn into long-lived open clusters, the backtracking of cluster sizes is of higher quality as the number of ejected stars is higher and the ejection happens over larger areas of the cluster \citep[][]{Pfalzner:2013}.

\subsubsection{Pre-expansion cluster size}

Using our simulation results as a starting point for backtracking, we find that the restriction to the unbound stars gives the best result in determining the pre-expansion cluster size. This can be seen clearly in Fig. \ref{fig:cut-off-trace-back} (top panel), where backtracked half-mass radius has been plotted against time. Backtracking the bound members provides no information, whereas using just unbound members fares much better. It recovers the half-mass radius ($r$\textsubscript{hm}) of the cluster at the time of gas expulsion with a relative error of $121.4_{-15.0}^{+16.3}$\% to the relative error of $298.9_{-46.7}^{+48.1}$\% obtained using bound members.  

It is equally important to include the unbound stars from a sufficiently large area.  Fig. \ref{fig:cut-off-trace-back} (bottom panel) shows a comparison of the backtracked half-mass radius determined by considering different areas for the member sampling. The horizontal lines show the derived pre-gas expulsion half-mass radii. It can be seen that the half-mass radius derived from the unbound stars sampled from a relatively small area (10 pc) results in a considerably larger error than those derived from including the unbound stars from larger areas. In relative error terms, the error decreases from $248.9_{-27.4}^{+41.6}$\% to $149.1_{-14.2}^{+16.1}$\% to finally, $121.4_{-15.0}^{+16.3}$\% as the search area around the cluster centre increases from 10 pc to 20 pc to 40 pc.
The actual size of the ideal backtracking area depends, among others, on the cluster's mass. Details on this dependence can be found in Pfalzner et al. (in preparation).

 Our simulations work with the idealised situation, where the search areas are uncontaminated by the presence of a population of foreground and background stars. In an actual application, extending the field increases the contamination by these foreground and background stars. A more significant fraction of contaminants yields a larger half-mass radius estimate and a shorter age estimate. As the ideal search radius increases as a function of cluster age, so do the errors due to the background population. However, the advent of \textit{Gaia} again improved the situation; nevertheless, it is still a point to consider in real applications. While \citet[][]{Rizzuto:2012} found ten years ago that the disc fractions in Upper Sco depend very much on cluster membership probability and distance to the cluster centre, nowadays, a search area of $>$ 100 pc is regarded as giving reliable data \citep[][]{Luhman2020}.

\subsubsection{Time of gas expulsion}

Backtracking can also be used to obtain information concerning the time when gas expulsion happened. Here the same rules apply as for determining the pre-gas expulsion size: restricting to unbound stars and including sufficiently large sampling areas improve the results. In the example shown in Fig. \ref{fig:cut-off-trace-back}, the simulated and the backtracked time of gas expulsion are shown as vertical lines. The backtracking of unbound members determines $t$\textsubscript{emb} to be 1.8 Myr, which is in excellent agreement with the actual value from the simulations (2 Myr, see Fig. \ref{fig:cut-off-trace-back} top panel). The relative error in gas expulsion time derived using unbound stars is $40\pm 4$\% which is much better than that derived using bound stars ($826_{-84}^{+45}$\%).
Moreover, including only the unbound particles within 10 pc is not advisable with its relative error of $88_{-32}^{+11}$\% in the recovery of $t$\textsubscript{emb}. The error is reduced to $63_{-8}^{+11}$\% when the search area increases to 20 pc. Although the results derived by including the unbound particles within 20 pc and 40 pc of the cluster's centre give nearly identical results for this example cluster (see Fig. \ref{fig:cut-off-trace-back} bottom panel), the relative error in the derived $t$\textsubscript{emb} decreases significantly to $40\pm4$\% when all the $N=1000$ simulations are considered for the 40 pc case. The derived gas expulsion times tend to underestimate the time of gas expulsion by a $32_{-6}^{+7}$\%. Given the general uncertainty of cluster ages, this can be considered a minimal error.  Again, it is the stars that underwent close encounters that are responsible for the derived too short times. 

\subsubsection{Further improvements}
We saw that using the unbound stars from a sufficiently large area gives the best backtracking results for the pre-gas expulsion half-mass radius. However, the value can still be a factor of two too large. One reason is that even some of the unbound stars have a relatively strong encounter before leaving the cluster (see Fig. \ref{fig:cut-off}). However, the main reason is that backtracking the unbound stars gives the half-mass radius of the unbound, not that of the entire cluster sample. The stars that become unbound are predominantly located at the outskirts of the cluster at the moment of gas expulsion. Therefore, backtracking them, one obtains a value that is larger than the complete half-mass radius. The actual pre-gas expulsion half-mass radius includes the unbound stars. However, simply multiplying the determined value by a factor of 0.5 recovers the half-mass radius in our case quite well. For our simulations, the empirical scaling factor has a value of $0.46_{-0.04}^{+0.06}$. There does not seem to be any correlation between the cluster mass and the scaling factor. Although the Spearman correlation coefficient is calculated to be $-0.0133$, the $p$-value for the hypothesis test of their correlation is found to be $0.48$ which is greater than the significance level $\alpha=0.05$. Hence, the null hypothesis that the cluster mass and the scaling factor are unrelated can not be rejected. To some degree, the actual correction value might depend on the star formation efficiency in the clusters, however, new sets of simulations with varying star formation efficiencies need to be analysed to establish the dependence. The gas dispersion timescale, on the other hand, should not affect the factor.

\begin{figure}[t]
    % \centering
    \includegraphics[width=0.5\textwidth]{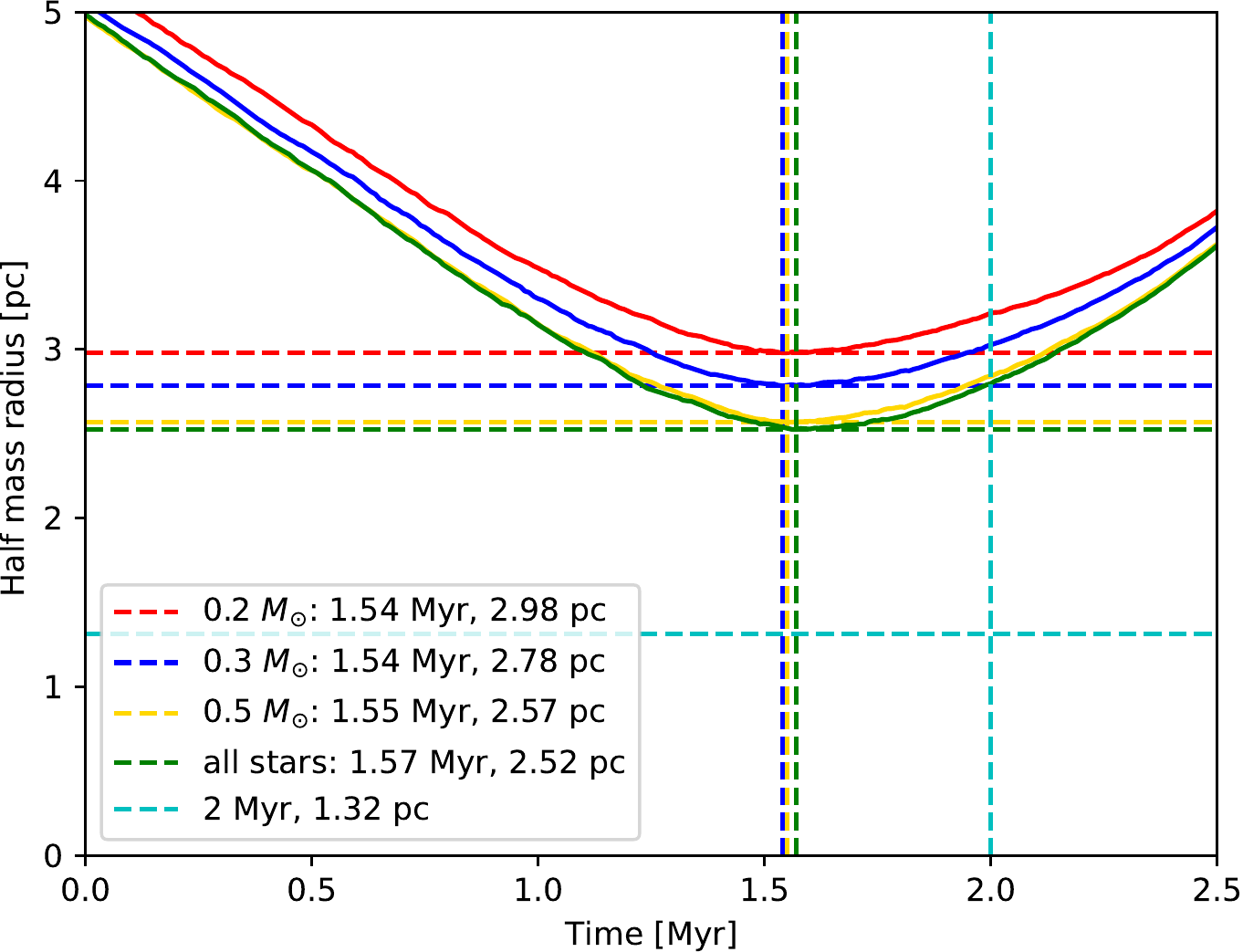}
        \includegraphics[width=0.5\textwidth]{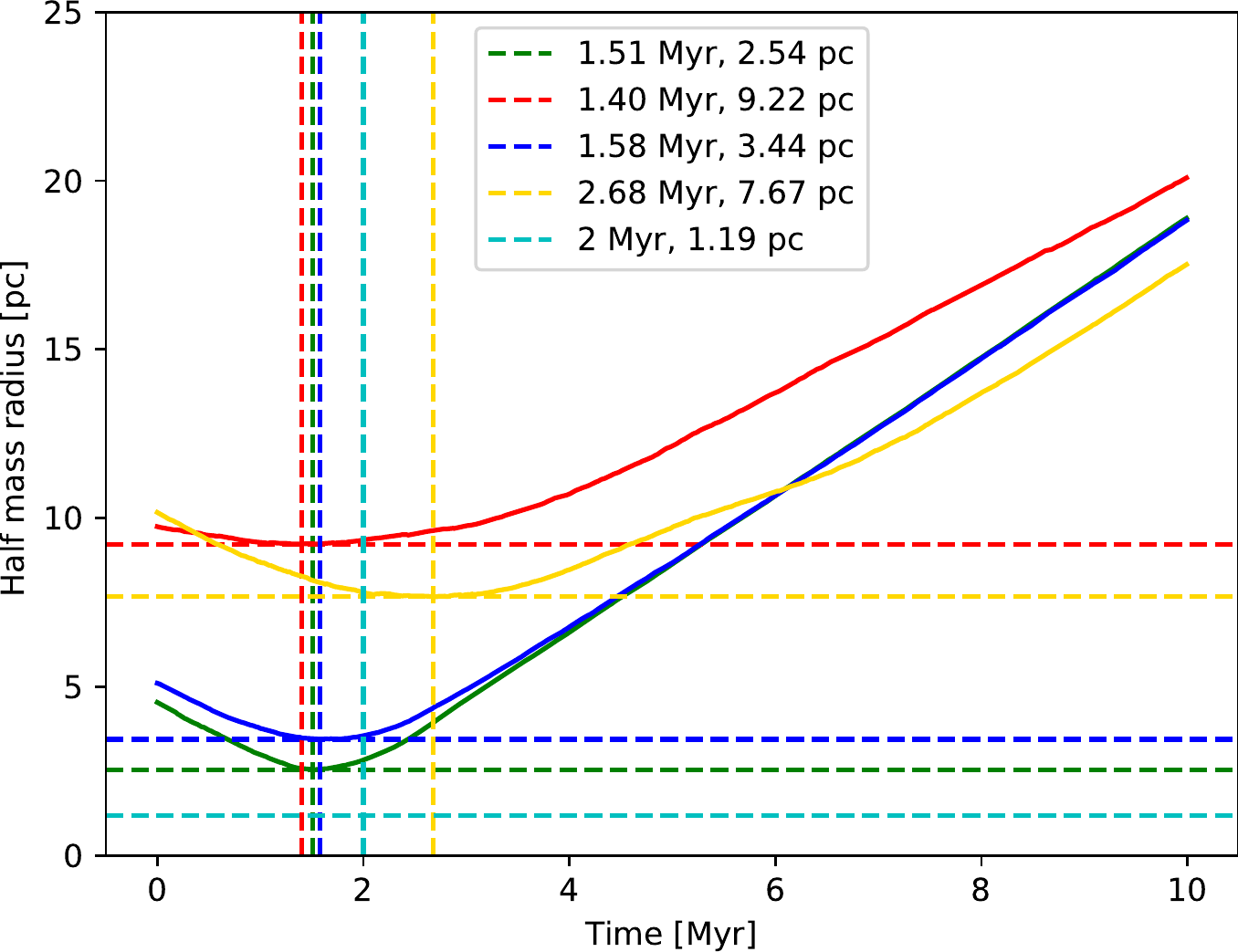}
    \caption{Backtracked half-mass radii for a simulation with 4000 stars, Top: calculated using actual masses (green), 0.2 $M_{\odot}$ (red), 0.3 $M_{\odot}$ (blue) and 0.5 $M_{\odot}$ (yellow). Bottom: calculated using exact velocity values (green), using $v_z=0$ (red), using velocities values with systematic errors as well as different levels of statistical uncertainty (blue: 0.27 km/s \& yellow: 1 km/s). The actual values of $t$\textsubscript{emb} and $r$\textsubscript{hm} at the time of gas expulsion from the simulation are shown in cyan.}
    \label{fig:mass_vz_hmr}
\end{figure}

\subsubsection{Mass of stars}
\label{subsubsec:mass}
When we determine bound and unbound stars in a cluster, the mass of the stars plays a role. However, in observations, the stellar classification is often known but not the actual mass of the stars. Especially for young clusters, there are large uncertainties between these two properties, and the assumption of different evolutionary models leads to significant differences. Here, we test to what extent this uncertainty in classification as bound or bound due to missing mass information influences backtracking. 

To mimic this problem, we assign the same mass to all stars, determine the bound and unbound stars and then perform the same backtracking procedure as before. Figure \ref{fig:mass_vz_hmr} (top) shows the result of backtracking with the fully known IMF (green) and with the assumption that all stars have the same mass ($M_s$ = 0.2 $M_{\odot}$, 0.3 $M_{\odot}$ and 0.5 $M_{\odot}$). It can be seen that not knowing the actual masses of the stars does not influence the derived time of gas expulsion. In all cases, it is too low.
The relative error for the derived $t$\textsubscript{emb} is $46_{-2}^{+3}$\% for the case of using actual stellar masses (green curve). Using the same stellar mass for all stars increases this error only marginally to $52_{-4}^{+3}$\%, $49_{-2}^{+3}$\%, and $47_{-2}^{+3}$\% for the case of $M_s$ = 0.2 $M_{\odot}$ (red), 0.3 $M_{\odot}$ (blue), and 0.5 $M_{\odot}$ (yellow) respectively.
The situation is different for the cluster size at the moment of gas expulsion. Here, assuming that all stars have the same mass leads to up to a factor of 1.2 larger sizes than using the actual stellar masses in the case shown in Fig. \ref{fig:mass_vz_hmr} (top). The smaller the assumed mass, the error is larger. The relative error for the derived $r$\textsubscript{hm} is $124.4_{-5.3}^{+8.5}$\% for the case of using actual stellar masses (green curve). This error increases to $130.6_{-7.0}^{+9.6}$\% when using stellar mass as 0.5 $M_{\odot}$ (yellow), to $155.6_{-9.4}^{+11.2}$\% for 0.3 $M_{\odot}$ (blue), and to $180.1_{-12.1}^{+15.0}$\% for 0.2 $M_{\odot}$ (red).\footnote{The distributions of sizes and gas expulsion times derived using different masses can be seen in Appendix \ref{Asec:mass}.}
We find that assuming all stars to have a mass of 0.5 $M_{\odot}$, which corresponds to the mean stellar mass in the cluster, is the best alternative to knowing the actual stellar masses.

\subsubsection{Velocity in the $z$ direction}
\label{subsubsec:vz}
%Another value that is not very well constrained in observations is the line-of-sight velocity. Again, we test the effect of uncertainty in the line-of-sight velocity on the backtracked radius values. To account for this problem, we modelled the extreme situation of zero information on $v_z$. The result is shown in Fig. \ref{fig:mass_vz_hmr} (bottom panel). The missing information of the z-component of the velocity leads to a nearly four times higher value than when $v_z$ is known. As the simulated half-mass radius is even smaller, the information of the former cluster size is highly unreliable when $v_z$ is unknown. However, even for unknown  $v_z$, the backtracked gas expulsion time remains basically the same. 

We also consider the effects of errors in the $v_z$ values on the backtracking in Fig. \ref{fig:mass_vz_hmr} (bottom). The velocity component along the $z$ axis, corresponding with close approximation to the radial velocity component, constitutes the main source of uncertainty in the total velocity vector \citep{Krolikowski2021}. As a starting point, we consider the effect induced by the existence of non-null proper motion uncertainties; the error on radial velocity is for the moment assumed to be null. \textit{Gaia} DR2 data have systematic uncertainties in the measurement of parallax and proper motions \citep{2018lindegren,Vasiliev2019}. 
% At a distance of $\sim 200$ pc, this systematic uncertainty translates to $\sim 0.06$ km/s. 0.13km/s
The 2D random error is considered to be of the order of 0.27 km/s, equivalent to the error in 2D proper motion (0.28 mas yr$^{-1}$) for sources with $G=17$ mag at a distance of 200 pc in \textit{Gaia} DR2. Using this error, blue curve is obtained for backtracked radii. The pre-expansion size is derived to be about 1.5 times the size obtained compared to the velocities having no error (green curve in Fig. \ref{fig:mass_vz_hmr}, bottom). The relative error distributions (with respect to the actual $r$\textsubscript{hm}) are determined for $r$\textsubscript{hm} obtained using velocities with no error (green) and using velocities with error (blue). The relative error in $r$\textsubscript{hm} goes from $124.4_{-5.3}^{+8.5}$\% for the green curve to $213.7_{-10.5}^{+12.7}$\% for the blue curve.
An accuracy improvement is seen for the value of the cluster's age at the time of gas expulsion. The relative error decreases from $46_{-2}^{+3}$\% for the green curve to $35_{-5}^{+4}$\% for the blue curve. However, this improvement is less due to recovering more information about the cluster's past, but more with a general move of the curve towards the right on the time axis with an increase in the standard deviation in random errors. 

The impact of radial velocity errors results in an even shorter estimate of the expansion timescale. \cite{Krolikowski2021} point out that the radial velocity (RV) uncertainty is roughly an order of magnitude larger than the reported projected proper motion uncertainty, even when collecting RV measurements from more precise catalogues than \textit{Gaia}.\cite{Ma2022} also point out that even with future \textit{Gaia} releases, the precision of RV would be $\sim 1$ km/s. The yellow curve in Fig. \ref{fig:mass_vz_hmr} (bottom) corresponds to the backtracked radii determined using the same systematic error but a random error of 1 km/s. This increases the relative error in $t$\textsubscript{emb} and $r$\textsubscript{hm} at the time of gas expulsion to $60_{-13.5}^{+8}$\% and $639.0_{-41.1}^{+35.9}$\% respectively.

Only 0.54\% of the sources with astrometric data have the RV measurements available in \textit{Gaia} DR2. For the extreme situation of zero information on $v_z$, the red curve in Fig. \ref{fig:mass_vz_hmr} (bottom) is obtained. The relative error for the determined size in this case is the highest of all previously discussed cases at $821.6_{-55.5}^{+47.6}$\% whereas the relative error in derived time of gas expulsion is $40_{-12}^{+10}$\%\footnote{The distributions of values of size and gas expulsion time obtained for all the cases discussed here can be seen in Appendix \ref{Asec:vz}}. In reality, for \textit{Gaia} DR2, the deviation from the actual parameter values will be somewhere between the cases of $v_z=0$ and the added systematic error along with statistical uncertainty.

\begin{figure*}[h]
   \includegraphics[width=0.32\textwidth]{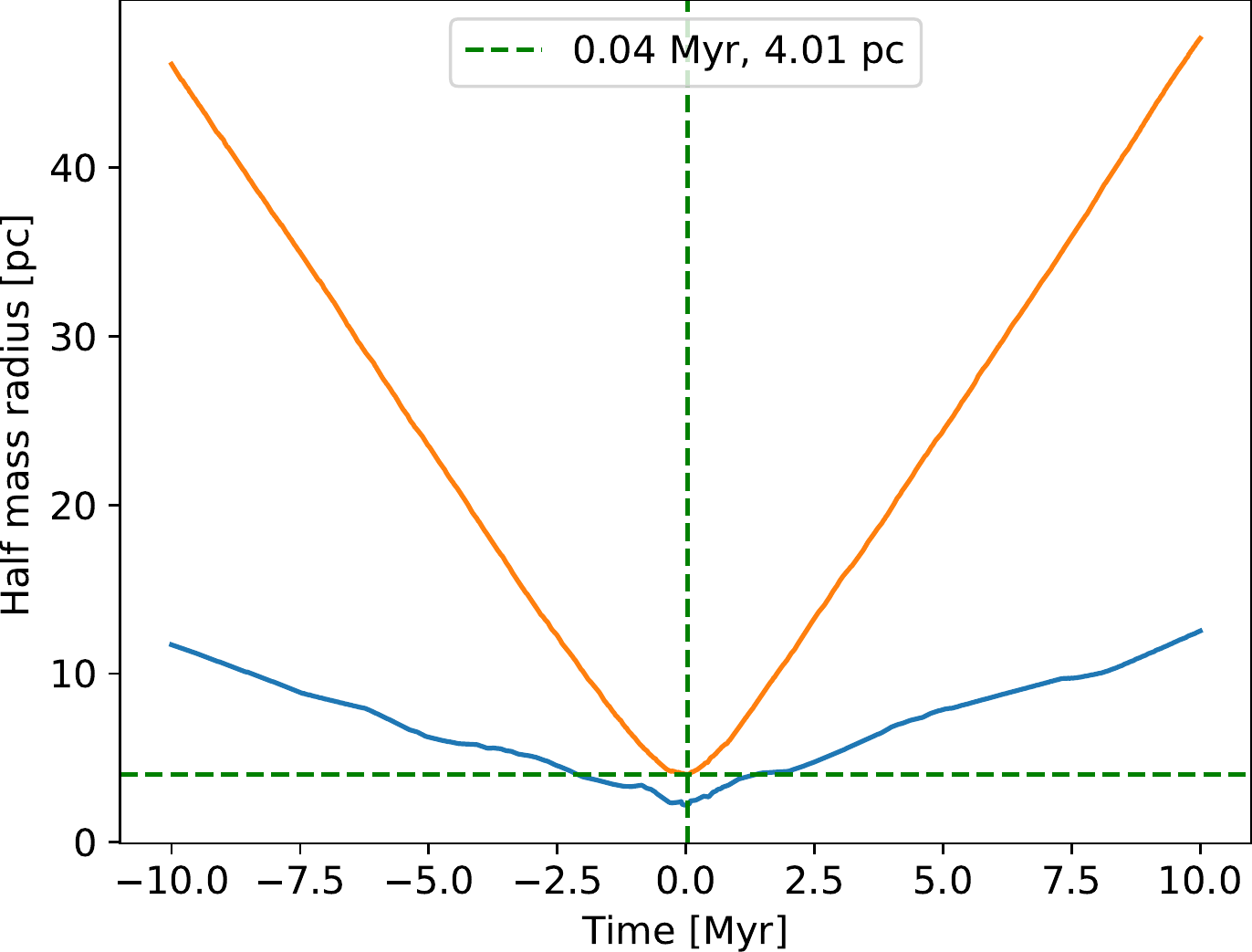}
\includegraphics[width=0.32\textwidth]{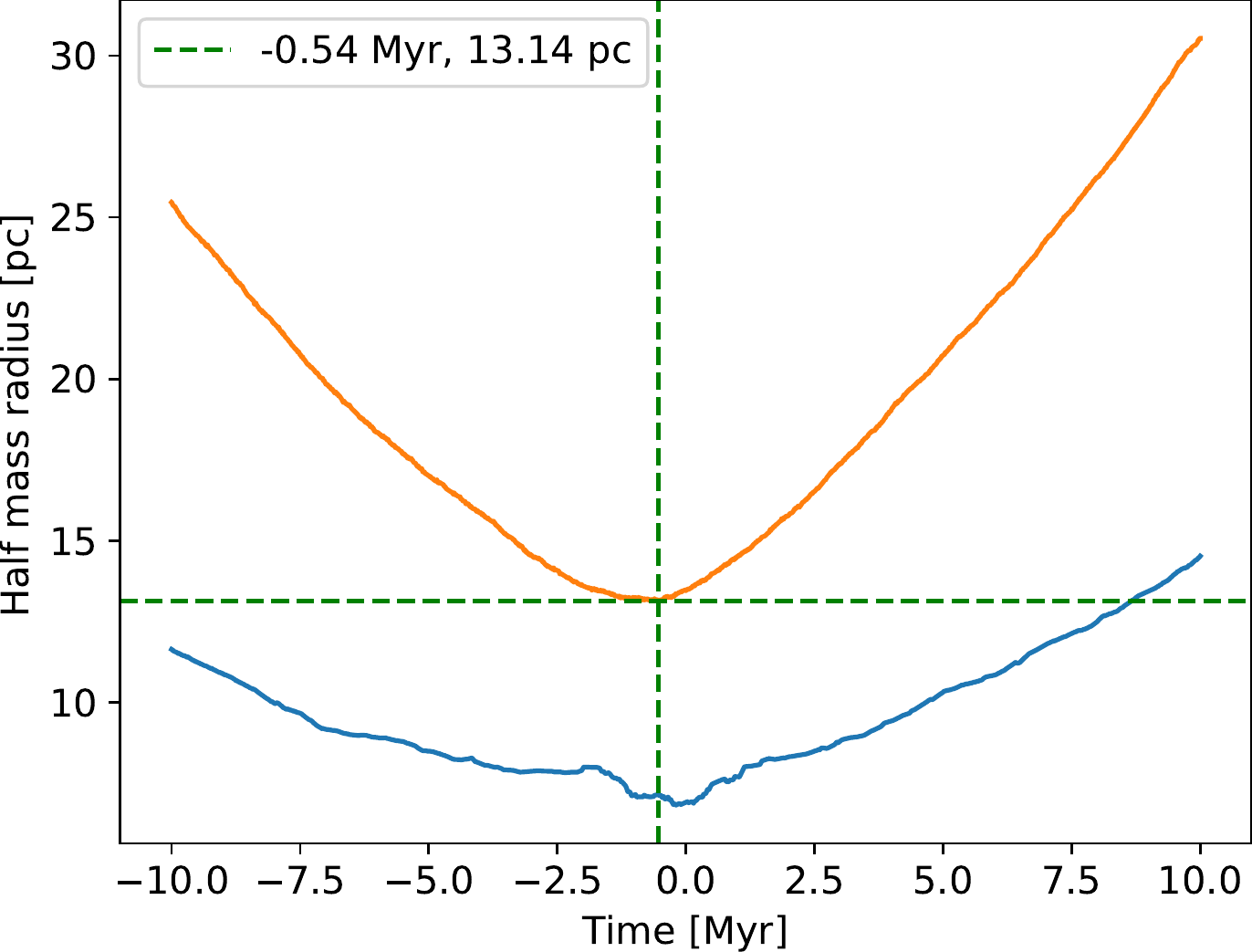}
    \includegraphics[width=0.32\textwidth]{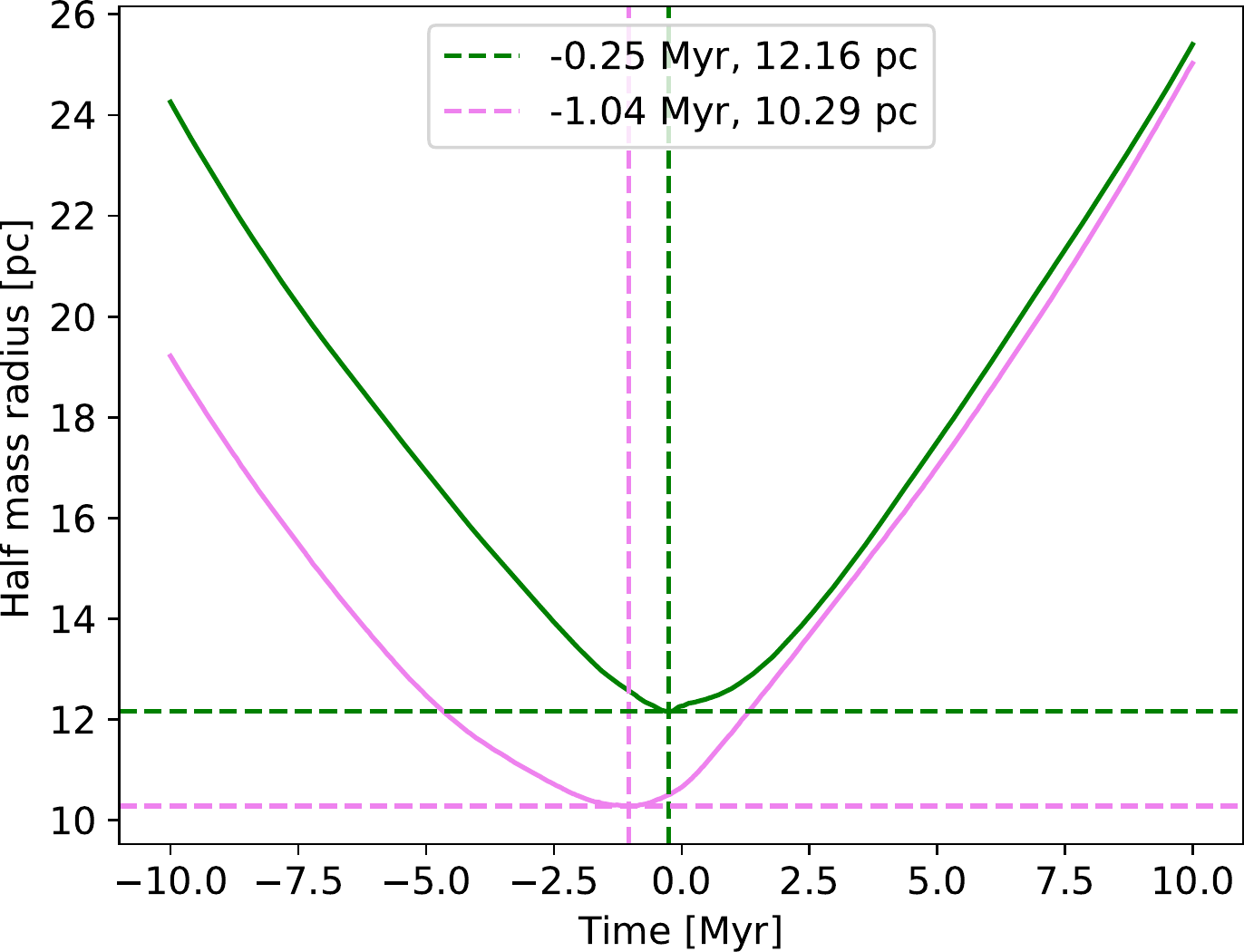}

\caption{Backtracked (and extrapolated) half-mass radii determined for bound (blue) and unbound (orange) stars 10 Myr into the past and into the future. The green dashed lines show the minima of the backtracked half-mass radius for unbound stars. Left: For NGC 6530 members. Middle: For Upper Sco members. Right: Backtracked (and extrapolated) half-mass radii determined for the unbound members of subclusters of Upper Sco.}
\label{fig:usco_bound}
\end{figure*}

\section{Application to observational data}

So far, we have dealt exclusively with the idealised situation that simulations provide. In the following, we want to show two examples of applying backtracking procedures to observed clusters. The aim is not so much the age and initial size determination of these specific clusters, but to show which additional problems can be expected in real applications. Therefore, we choose two clusters that differ considerably in age and geometry. When referring to the age of the cluster, we quote the time elapsed since the gas started to be expelled and refer to the cluster age as the median age of all the stars in the cluster. This differs from the time elapsed since the molecular cloud started producing stars \citep[][]{Pecaut:2016,Kim:2021,Fujii:2021}.

\subsection{NGC 6530}

We first apply the before-described backtracking method to NGC 6530, which is a young cluster within Lagoon Nebula. Its age has been estimated to be 1--2.3 Myr \citep{ngc_age2005,mayne2007,Bell:2013} and its distance to be \textbf{$1326^{+77}_{-69}$ pc} \citep{Wright2019,Damiani2019}. We use the catalogue of members provided by \cite{Wright2019}, who use GES spectroscopy, \textit{Gaia} DR2 astrometry, and ancillary membership information from X-ray, infrared, and H$\alpha$ surveys to compile the said catalogue. 691 of these cluster members have \textit{Gaia} DR2 data and have been used in the following analyses.
We assume that all the stars have a mass of 0.5 M$_{\odot}$. Using the radial velocity for individual sources when available and assuming it to be equal to the bulk radial velocity of the cluster when not, 3D positions and velocities of the stars are calculated in the standard right-handed Cartesian Galactic frame using the conversion equations prescribed by the \textit{Gaia} DR2 documentation. 
These are then used to determine the bound and unbound members of the cluster. 

For backtracking the stars' trajectories, we backtrack the positions in the plane of the sky using the velocities along $\alpha$ and $\delta$. Radial velocity is used to backtrack along the line-of-sight and change the distance of the stars which is assumed to be the same for all stars at the present time (1326 pc). Although individual distances are available for all the stars \citep{bailerjones2018}, the uncertainty is extremely high (fractional uncertainty is $0.20_{-0.09}^{+0.43}$ as compared to $0.02\pm0.01$ for the distance data-set of member stars of Upper Sco in Sec.\ref{subsec:upper sco}) and leads to very high half-mass radius along with loss of most information about the cluster. The calculated coordinates are then converted to the Cartesian coordinates to calculate the half-mass radii. The result of this procedure is shown in Fig. \ref{fig:usco_bound} (left panel). However, for considering the uncertainty in astrometry of the member stars, we run $1000$ Monte Carlo simulations, that is to say repeat the entire procedure while varying astrometric information in a random, normal manner according to the uncertainties associated with each \textit{Gaia} DR2 source's parameters. For the distance value for all the stars, the uncertainty is taken as 73 pc \citep{Wright2019}. The results of these simulations are fitted with a Gaussian to obtain the parameters of the cluster along with their errors. Hence, we find the gas expulsion to have happened $0.03\pm 0.03$ Myr ago and the size of the cluster at the time of gas expulsion is found to be $4.16\pm 0.23$  pc.
% Since most clusters expel their residual gas by 1--2 Myr \citep{Pfalzner2019}, an upper limit of \textbf{1.03--2.03 Myr} is put on the age of NGC 6530.
This agrees well with the current age estimate of the cluster. However, the half-mass radius might be underestimated by the assumption of a fixed distance of the stars. A more realistic estimate might be obtained by multiplying it by a factor $\sqrt{3/2}$, which would yield a limit of 5.09 pc on the cluster size at the time of gas expulsion.

Despite obtaining a reasonable fit, the reservations pointed out in Section 4.2.5 also hold here. The median uncertainty in proper motion amount to 2 km/s \citep[][]{Wright2019}. Any uncertainty added to the true velocity acts to reduce the best fit. This uncertainty is the most problematic issue in applying the backtracking method for determining the age of NGC 6530.

\subsection{Upper Scorpius}
\label{subsec:upper sco}
Upper Sco is a sub-group of Sco-Cen that has been widely studied with the \textit{Gaia} data, identifying the cluster's members \citep{Galli2018,Wilkinson2018,Luhman2020,Damiani2018,chronostar2,Squicciarini2021,kerr_etal2021} 
and an isochronal age of around 10 Myr has been recently accepted \citep{feiden16,david19,Luhman2020,sullivan_kraus21}. 
%Thus, this cluster is, in principle, suited for applying the backtracking analysis that we have deduced from the simulation data till now. 
We test the quality of the backtracking for clusters with a more complex morphology using Upper Sco as an example. 
We use the list of members compiled by \cite{Luhman2020} using optical and IR spectra to confirm the stars' youth while parallax and proper motion offsets to get the kinematic criteria for these candidates. The list contains 1761 member candidates, 1682 of which have \textit{Gaia} DR2 data available and have been used in the following analyses. We apply the same method described for NGC 6530 with the exception of considering individual distances for the stars in this case as the uncertainty in distance is much lower.
% , assuming a median distance of 145.58 pc \citep{bailerjones2018}. 

% shown in Fig. \ref{fig:usco_bound} (a). The red horizontal and vertical lines show the velocity and distance cutoffs determined for $N=4000$ cluster shown in Fig. \ref{fig:cutoff_boxplot} while the green line shows the analytical escape velocity cutoff with Plummer radius taken to be equal to the half-mass radius of the $N=4000$ simulated cluster (1.3 pc) for reference. Considering the bound and unbound membership determined using 3D positions, velocities and the assumed 0.5 M$_{\odot}$ mass to be correct, we find a rate of correct identification of around 93\% for the analytical cutoff and around 45\% for the distance-velocity cutoffs. The rate of correct identification increases to $\sim 94 \%$ if the stars satisfying either the distance cut-off or the velocity cut-off are classified as unbound.  
Despite its complex morphology, we first work with the assumption that Upper Sco was a centrally condensed spherical structure in the past. In this case, we find that the cluster went through gas expulsion 0.54 Myr ago and had a half-mass radius of 13.14 pc at this time as shown in Fig. \ref{fig:usco_bound} (middle panel). However, the Monte Carlo simulations for error propagation estimation provide the gas expulsion time to be $0.80\pm0.21$ Myr ago while the cluster size is found to be $13.11\pm0.11$ pc.
% Considering the time taken by the cluster to expel its residual gas to be 1--2 Myr, this would put an upper limit of {\bf1.80--2.80 Myr} on the age of the Upper Sco association. 

This value agrees with other backtracking results for Upper Sco. For example, \cite{chronostar2} determine the kinematic age of the population in the Upper Sco region as $4 \pm 4$ Myr, whereas \cite{Squicciarini2021} find 8 subclusters with kinematic ages varying from $0.0 \pm 0.1$ Myr to $3.8 \pm 0.4$ Myr. However, this cluster age deviates considerably from that of 10 Myr obtained by applying corrections, for undetected binaries \citep{sullivan_kraus21} or strong magnetic fields impeding convection in low-mass stars \citep{feiden16,david19}, to the isochronal age determination of Upper Sco. One possible explanation for this discrepancy would be that the backtracking yields the time elapsed since gas was expelled and refers to the age of the youngest stars in the association. Taking into account a star formation history lasting 6-7 Myr, most stars might be about 11 Myr old and the median age of the association $\approx$ 7 Myr. These values are more similar to the ones obtained through stellar evolution models. 

%  prevent one from using the backtracking method: as the authors say, US is highly substructured; the literature studies mentioned in the text have shown that the SFH of the association  Moreover, the current size of US is too large to be the result of the expansion of a single structure given its current velocity dispersion 

 Additional complications arise from Upper Sco, unlike NGC 6530, being highly substructured \citep{kerr_etal2021,Squicciarini2021}. Likely, star formation did not happen as a single burst, but was rather characterised by several formation episodes \citep{Galli2018}. Thus, the assumption of a centrally condensed spherical structure in the past is oversimplifying the situation. Hence, we try to improve our analysis by considering Upper Sco to consist of subclusters. A density distribution of the cluster members on the plane of the sky at the present time is plotted (see Appendix \ref{Asec:upper sco} for more details and plots). Two dense areas seem to emerge and we consider two rectangles in these areas. The members' positions are traced back using the same method as described above. When a member star enters one of the said rectangles, it is assigned to the corresponding subcluster. After the assignment of subcluster membership using this simplified method, the backtracked and extrapolated half-mass radii are determined using unbound stars for both subclusters. The result is shown in Fig. \ref{fig:usco_bound} (right). To determine the errors, the Monte Carlo simulations are used which provide the time of gas expulsion in the two subclusters as $-1.09\pm0.29$ Myr and $-0.25\pm0.17$ Myr ago respectively. Similarly, the half-mass radii at the time of gas expulsion is found to be $10.15\pm0.20$ pc and $12.10\pm0.23$ pc. Various characterisations of the subclusters are summarised in Table \ref{Afig:uppersco}. There is a slight improvement in the determination of the size and time of gas expulsion when considering Upper Sco to have subclusters rather than being one coeval population. However, it must be reiterated that ours is a simplified method. More robust clustering methods can be used in the future to get better results on the subcluster membership and hence, their parameters. For example, \cite{kerr_etal2021} use HDBSCAN clustering algorithm on \textit{Gaia} DR2 data and find 9 subclusters in the Upper Sco region. Two of these (Group H and Group I) have more than 100 members. We analyse these subclusters and find the time of gas expulsion and their sizes at that time. According to our results, gas expulsion in Group H happened $3.40\pm 0.42$ Myr ago and its half-mass radius was $3.96\pm0.215$ pc at the time. For Group I, the gas expulsion happened $0.78\pm0.91$ Myr ago and its size was $3.73\pm0.37$ pc. The age found by \cite{kerr_etal2021}, using \textit{Gaia} DR2's photometric data, for the groups is $10.2 \pm 0.7$ Myr and $5.7\pm0.4$ Myr respectively. So, even though there is an improvement in the age and size estimates when using a more robust clustering algorithm, the kinematic age estimates still show considerable deviation from the photometric estimates. Availability of accurate radial velocities and distances for the member candidates to use in the subclustering analysis in future would improve the situation further.

% {\bf Likely applying backtracing to each subgroup individually would further improve the result. }

\section{Discussion}

The improvement in the cluster size, when considering subclusters, already shows that backtracking is more complex for substructured clusters like Upper Sco. Thus, the less substructured a cluster is, the more straightforward the backtracking. The substructured clusters require backtracking to multiple centres, which is the more complex the more subcluster centres exist. 

% While Fig. \ref{fig:usco_bound} (right panel) shows a way to improve the derived pre-gas expulsion size by assuming that Upper Sco consists of subclusters, gas expulsion time determination seems relatively unaffected. 
Another potential difficulty could be the presence of multiple differently aged populations in the Upper Sco region leading to the miscalculation of the cluster's age \citep{Wright2018,chronostar2,Squicciarini2021}. However, this would require large subgroups to be well over 15 Myr to introduce such a substantial error. This seems unlikely as an explanation.
We suspect that the real reason is a different one. The arguments based on kinematic analysis of a cluster for its history can not be considered on their own due to the significant errors in radial velocity and its unavailability for most stars in \textit{Gaia}. Large uncertainty in the velocities of the stars can lead to a significant loss of information about the past of the cluster (see Fig. \ref{fig:mass_vz_hmr}, bottom panel). This might be the reason for underestimating the cluster age and overestimating the size at the time of gas expulsion. Furthermore, the assumptions in the backtracking analysis are numerous. The exact masses of the stars are unknown, so the distinction between bound and unbound stars could be highly inaccurate when combined with astrometric uncertainties and incomplete or inaccurate membership of the cluster.
In conclusion, the determination of a much younger age, of the Upper Sco region, by kinematic analysis than the more accurate isochronal determination could be affected by multiple, differently aged and kinematically distinct populations; however, precise radial velocity measurements are needed to rule out the possibility that the discrepancy in age determination is due to astrometric errors. 

\section{Summary and conclusion}

Young star clusters ($<$ 10 Myr) are highly dynamical entities. Therefore, observations provide only snapshots of this highly dynamic cluster evolution sequence. Nevertheless, in light of the unprecedented precision of \textit{Gaia} position and velocity data, it should be possible to obtain information about a young cluster's past using backtracking techniques. In this work, we used simulations of the cluster dynamics as an idealised version to suggest how to optimise the backtracking method. Under ideal observational conditions, the following statements should hold:

\begin{itemize}
    \item For backtracking to be successful, it is essential to distinguish between bound and unbound cluster members.
    Under ideal conditions, backtracking the unbound members exclusively, the time of gas expulsion can be determined with only a 32\% error. However, the quality of the backtracking depends on the number of cluster stars, with the best results obtained for clusters containing a few thousand stars.
    
    \item While still the best result, the sizes backtracked from unbound members are about a factor of two larger than the actual value. However, this error is systematic and reflects that unbound members are primarily located at the cluster outskirts at the time of gas expulsion. Thus, applying a correction factor of 0.46 approximates the actual value very well. 
    
    \item For obtaining this accuracy, it is essential to determine all the unbound members to $>$ 20 -- 40 pc from the cluster centre.
    
    \item The classification of bound and unbound stars based on the direction of their velocity vectors, or ad hoc distance or velocity cutoffs is highly error-prone. We provide analytical cutoffs based on the escape velocity and the number of cluster members with a success rate of 96\% -- 97\% for distinguishing between bound and unbound stars. 
    
    \item Runaway and walkaway stars are less suitable to determine past cluster properties because of their low number and their production by dynamical ejection. Ejection traces only past locations of high stellar density regions but not actual cluster sizes or the time of gas expulsion.
    
\end{itemize}    

Uncertainty in membership and stellar properties provide additional challenges. Modelling these uncertainties, we find that the lack of information about the line-of-sight velocity can severely affect the determination of the pre-expansion size of the cluster. Nevertheless, the time of gas expulsion can still be estimated with an error of $40\% - 60\%$ due to the unavailability of radial velocities and uncertainty in the value even when available. The uncertainty in the mass of the members seems to affect the results much less. Similarly, larger search areas often struggle with higher false-positive and -negative rates in membership. Applying our results to observational data, the method works reasonably for centrally concentrated clusters, but less for very substructured clusters like Upper Sco. For such substructured clusters, backtracking to the individual subcluster centres would be the next step to pursue.

In summary, restricting backtracking to the unbound stars allows deducing the times of gas expulsion and the pre-expansion cluster size values with relatively high accuracy. Analysing a large number of clusters with the presented method will allow drawing valuable conclusions about the clustered star formation process in the future.   

%
%\begin{figure*}[t]
%    \centering
%    \includegraphics[width=0.5\textwidth]{NEW/mass uncertainty.png}
%    \includegraphics[width=0.28\textwidth]{NEW/mass_zoom.pdf}
%        \includegraphics[width=0.5\textwidth]{NEW/limiting_mag_edit.png}
%    \includegraphics[width=0.33\textwidth]{NEW/lim_mag_zoom.png}
%    \includegraphics[width=0.33\textwidth]{2d_vs_3d_mean_hmr.png}
%    \caption{Top:Backtracked half-mass radius for unbound stars in an $N=4000$ cluster using the actual masses (in green), using stellar mass equal to $0.2 M_{\odot}$ (in red), $0.3 M_{\odot}$ (in blue) and $0.5 M_{\odot}$ (in yellow) for all stars. The actual values of $t_{emb}$ and $r_{hm,ge}$ from the simulations are shown in cyan. Zoom of the shaded area shown. Bottom:Backtracked half-mass radius for unbound stars in an $N=4000$ cluster using all the stars (in yellow), using only stars with masses greater than $0.5 M_{\odot}$ (in green), $1 M_{\odot}$ (in blue) and $2 M_{\odot}$ (in red). The actual values of $t_{emb}$ and $r_{hm,ge}$ from the simulations are shown in cyan. Zoom of shaded area shown. 2d vs 3d backtracing- Yellow,  Cyan:  3d backtrace and actual simulation values.   Green,  Blue:  2d backtrace and actual simulation values for time of gas expulsion and half mass radius}
    \label{fig:Mgt0.5}
    \label{fig:n4000_mass_hmr}
%\end{figure*}

% \begin{figure}[H]
%     % \centering
%     \includegraphics[width=0.5\textwidth]{hmr_comparison_mean_time.png}
%     \includegraphics[width=0.5\textwidth]{hmr_comparison_best_time.png}
%     \caption{Comparison of different boundaries of observed area: mean of simulations (left), best fit (right)N=1000}
%     \label{}
% \end{figure}

% \begin{figure}[H]
%     % \centering
%     \includegraphics[width=0.5\textwidth]{cutoff_hmr_1092.png}
%     \includegraphics[width=0.5\textwidth]{cutoff_hmr_1092_zoom.png}
%     \caption{Comparison of half mass radius calculated by backtracing unbound stars [1092: worst fit]: unbound members determined by energy calculation vs distance cutoff vs distance and energy cutoffN=1000}
%     \label{}
% \end{figure}

% \begin{figure}[h]
% \includegraphics[width=0.48\textwidth]{backtracing.pdf}
% \caption{2D projections: Snapshot of an example of our set of simulations a) at t=2 Myr and b) at 10 Myr. Backtracking from the results at 10 Myr to 2 Myr for (c) all stars and (d) only for the unbound stars. The bound stars are indicated in blue and the unbound stars in red. For each star the velocity vector is shown. A film of the cluster dynamics and the backtracking can be found at \textcolor{red}{eventually add link.}}
% \label{fig:2dbacktracking}
% \end{figure}

\begin{acknowledgements}
We thank the referee for a very detailed report that made this article
significantly better.
This work has made use of data from the European Space Agency (ESA) mission
{\it Gaia} (\url{https://www.cosmos.esa.int/gaia}), processed by the {\it Gaia}
Data Processing and Analysis Consortium (DPAC,
\url{https://www.cosmos.esa.int/web/gaia/dpac/consortium}). Funding for the DPAC
has been provided by national institutions, in particular the institutions
participating in the {\it Gaia} Multilateral Agreement.
\end{acknowledgements}

\bibliographystyle{aa} % style aa.bst
\bibliography{references} % your references Yourfile.bib

\begin{appendix}

\section{Mass of stars}
\label{Asec:mass}

\begin{figure}[h]
    % \centering
        \includegraphics[width=0.45\textwidth]{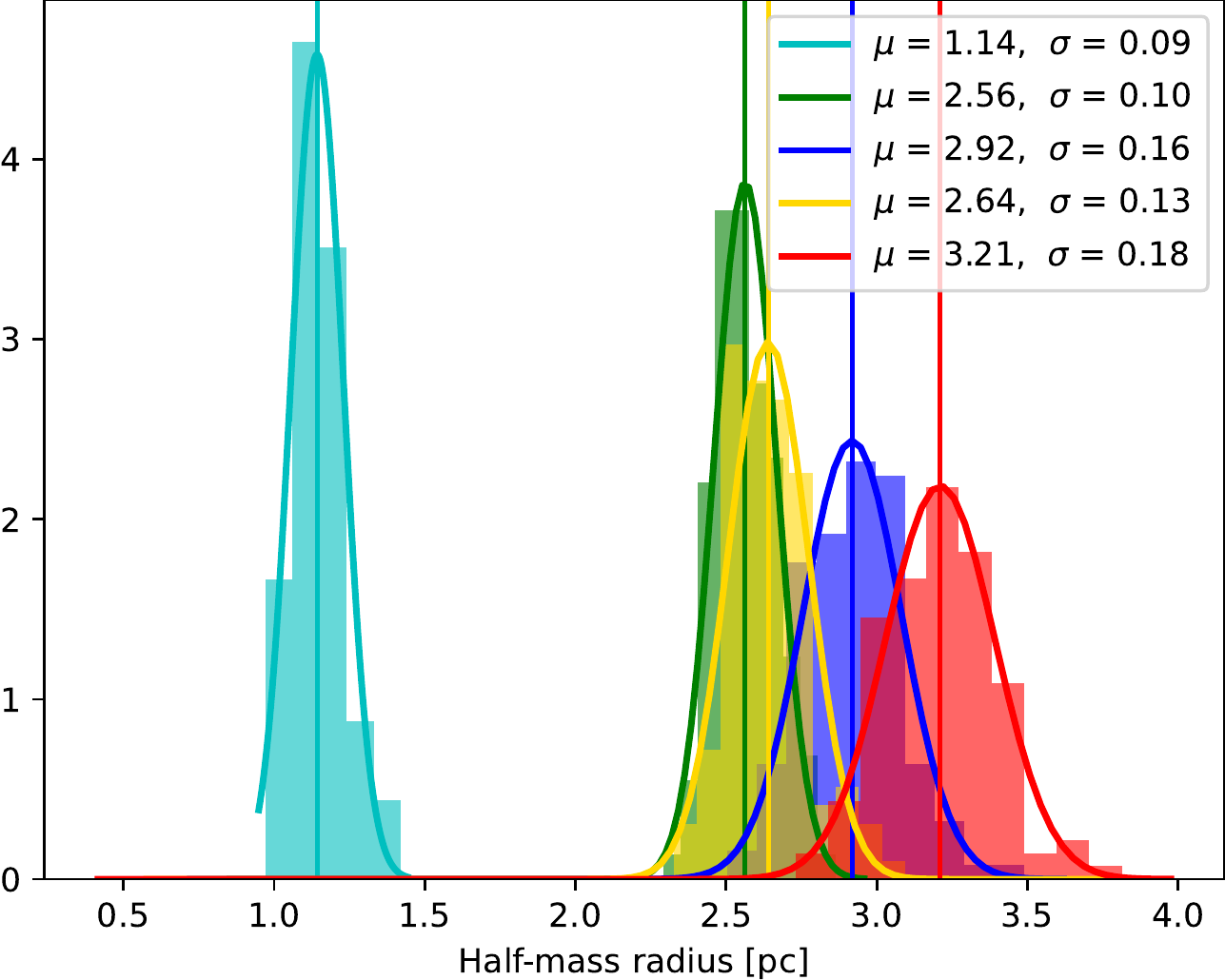}
    \includegraphics[width=0.45\textwidth]{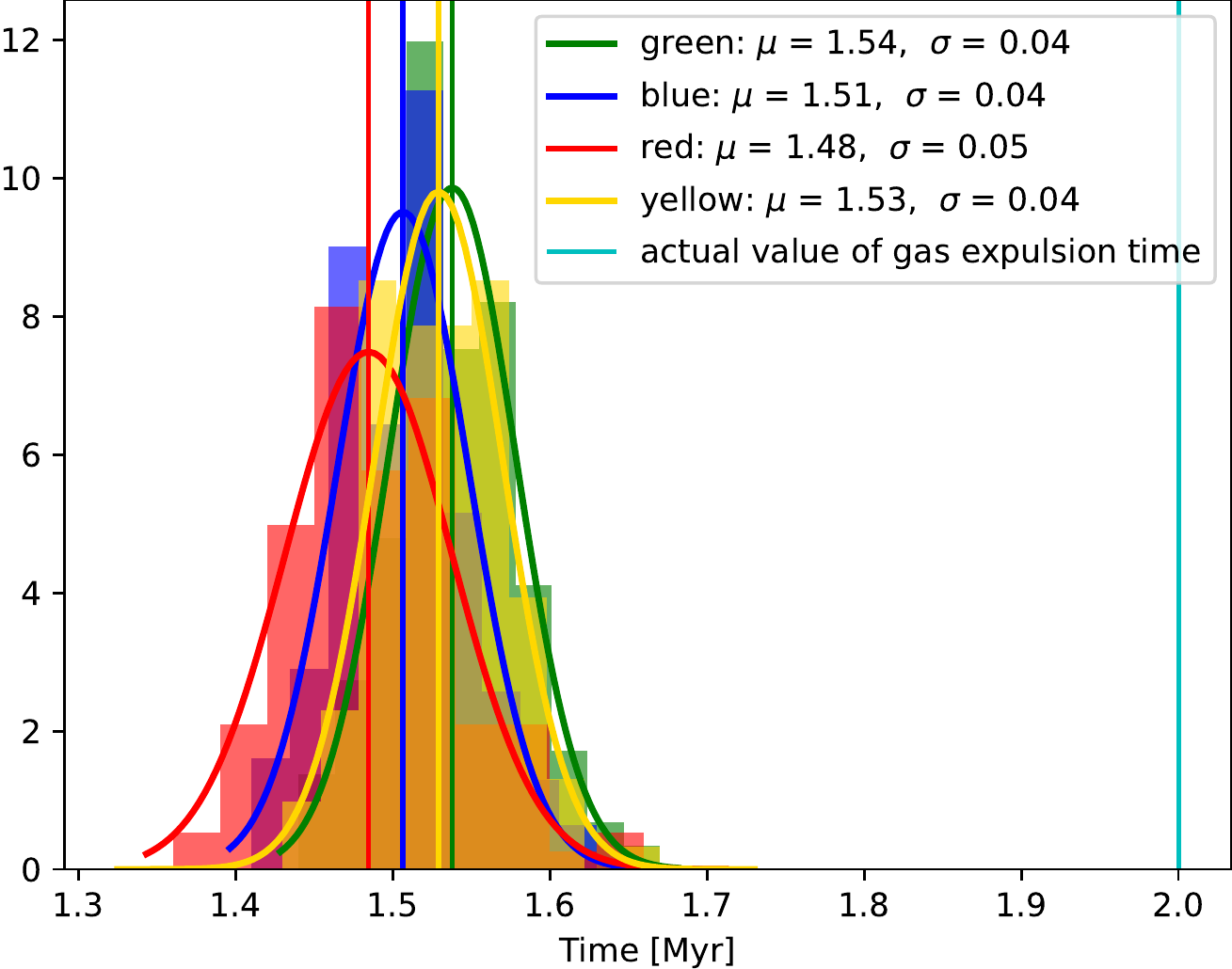}
    \caption{Distributions of the backtracked half-mass radii (top) and the time of gas expulsion (bottom) obtained using actual masses (green), 0.2 $M_{\odot}$ (red), 0.3 $M_{\odot}$ (blue) and 0.5 $M_{\odot}$ (yellow). The actual values of $r$\textsubscript{hm} at the time of gas expulsion (as a distribution) and $t$\textsubscript{emb} from all the simulations (of $N$=4000 clusters) are shown in cyan.}
    \label{Afig:mass_hmr}
\end{figure}

We discussed how the unavailability of the mass of stars in observations affects the determination of gas expulsion time and cluster size at the time of gas expulsion using backtracking analysis. Here, we provide the distributions of the derived sizes and gas expulsion time (Fig. \ref{Afig:mass_hmr}) for all the cases discussed in Sec. \ref{subsubsec:mass}.

\section{Velocity in the $z$ direction}
\label{Asec:vz}
\begin{figure}[h]
    % \centering
       \includegraphics[width=0.45\textwidth]{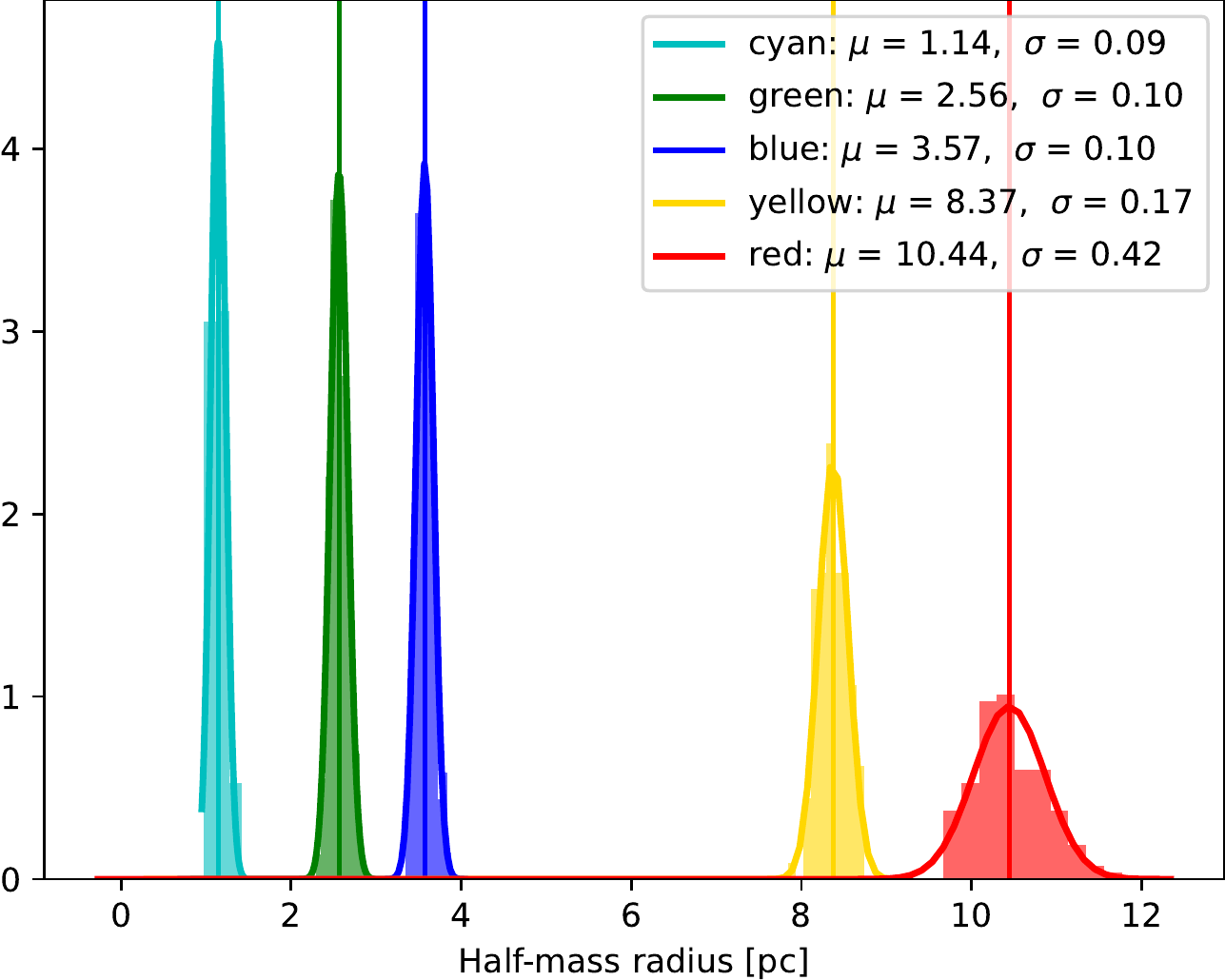}
    \includegraphics[width=0.45\textwidth]{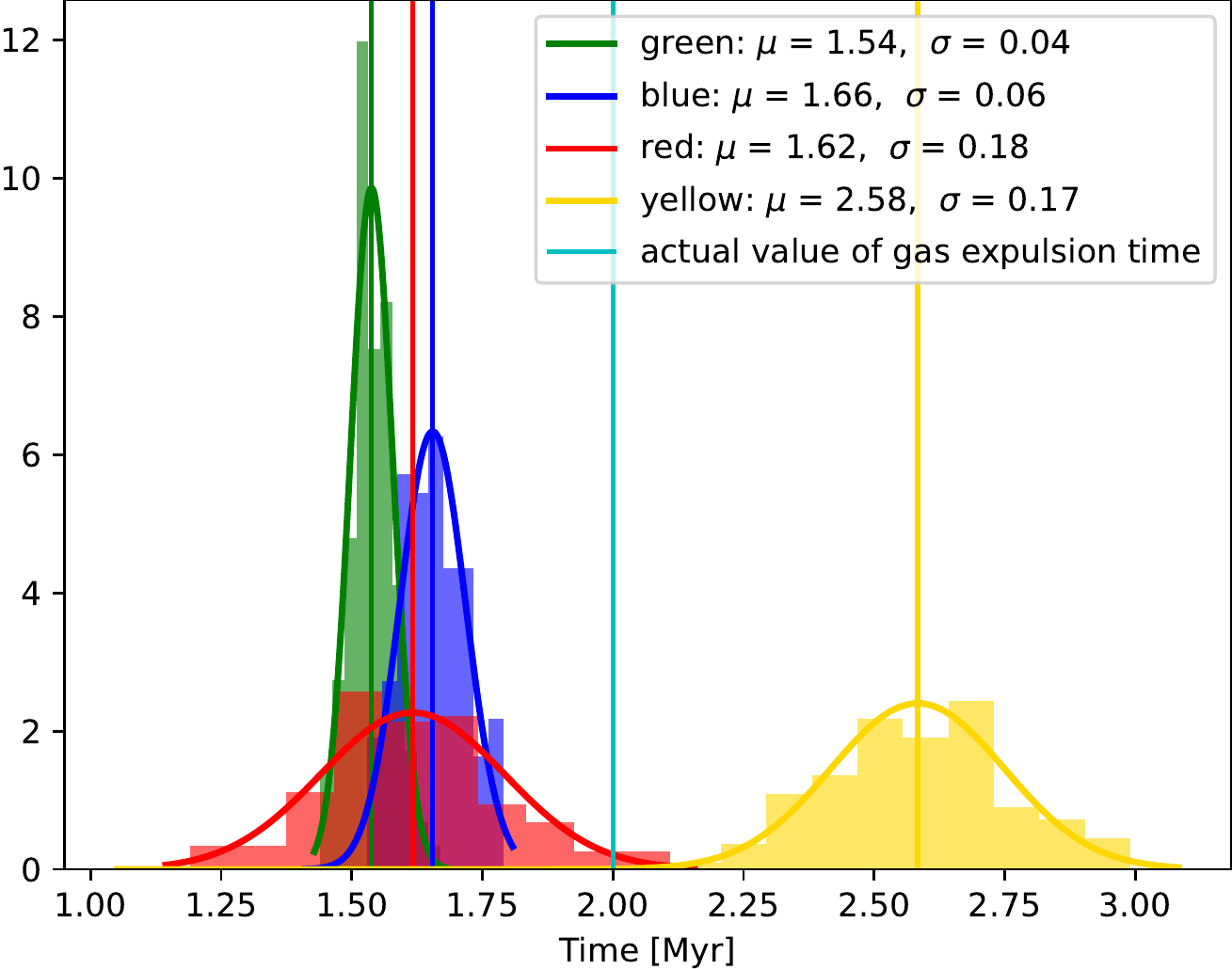}
    \caption{Distributions of the backtracked half-mass radii (top) and the time of gas expulsion (bottom) obtained using exact velocity values (green), using $v_z=0$ (red), using velocities values with systematic errors as well as different levels of statistical uncertainty (blue: 0.27 km/s \& yellow: 1 km/s). The actual values of $r$\textsubscript{hm} at the time of gas expulsion (as a distribution) and $t$\textsubscript{emb} from all the simulations (of $N$=4000 clusters) are shown in cyan.}
    \label{Afig:vz_hmr}
\end{figure}
Similarly, we provide the distributions of the derived sizes and gas expulsion time for all the cases in Sec. \ref{subsubsec:vz} to supplement the discussion of the effects of errors in the $v_z$ values on the backtracking analysis and derived parameters (Fig. \ref{Afig:vz_hmr}).

\section{Upper Sco subclusters}
\label{Asec:upper sco}
\begin{figure}[h]
    % \centering
    \includegraphics[width=0.5\textwidth]{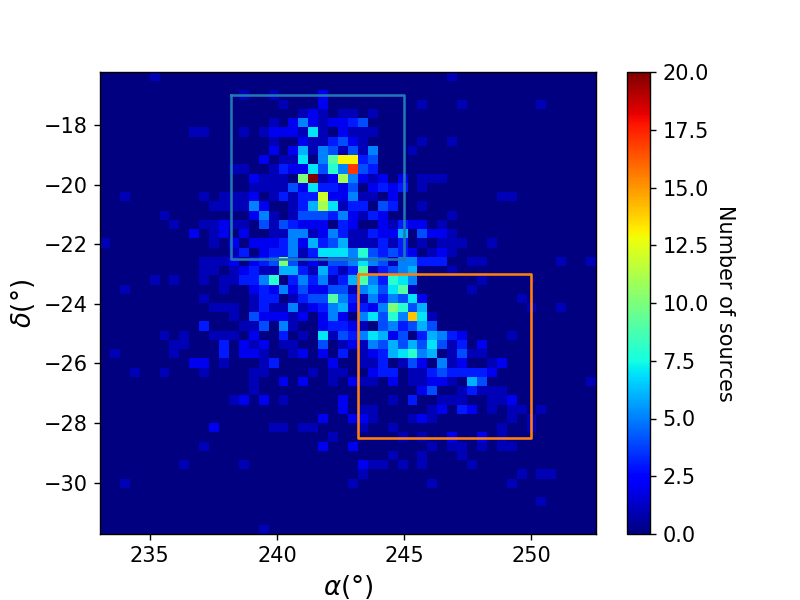}
    \includegraphics[width=0.5\textwidth]{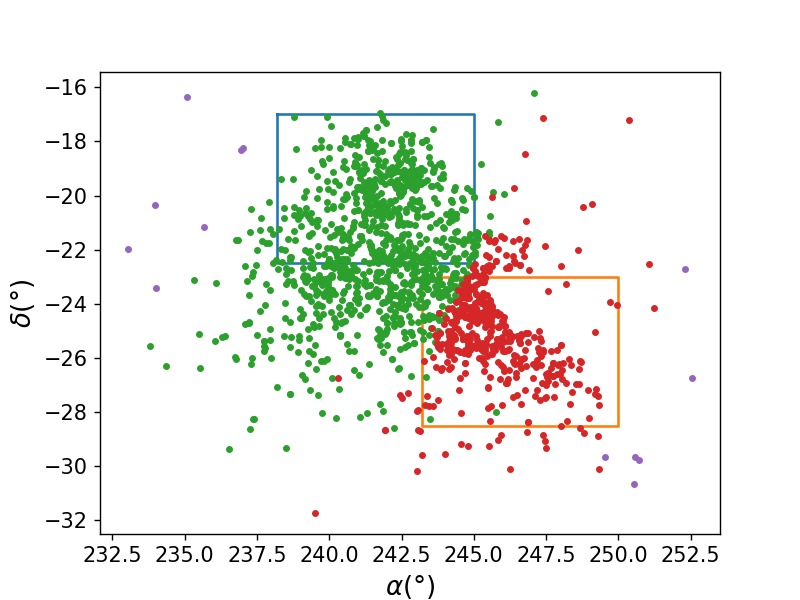}
    \caption{Density distribution (top) and scatter plot (bottom) of the Upper Sco members at the present time. The two rectangles show the area selected for the clustering process. Green and red points in the bottom plot show the members of Group 1 and Group 2, respectively. Purple points are the ones which were not assigned to any group. }
    \label{Afig:uppersco}
\end{figure}

The density distribution of the Upper Sco members is shown in Fig. \ref{Afig:uppersco} (top) along with the rectangles showing the subcluster areas used for the subcluster membership assignment. Figure \ref{Afig:uppersco} (bottom) shows the scatter plot of the member stars with the same rectangles and the members of the two subclusters in red and green. The purple points represent the few members which did not enter any of the rectangles in the 10 Myr up to which the positions were backtracked and hence, are not assigned to any subcluster. Furthermore, Table \ref{tab:subgroups} provides characteristic information about the subclusters identified in this work as well as about Group H and I from \cite{kerr_etal2021}.  

\begin{table}[b!]
    % \centering
    \caption{Information about the subclusters identified in this work (ID: 1,2) and the groups from \cite{kerr_etal2021} (ID: H, I).}
    \begin{tabular}{cccccc}
    \hline
    \hline
        ID & N & RA & Dec  & $t_K$ & $r$\textsubscript{hm}\\
        &&[deg]&[deg]&[Myr]&[pc]\\
        \hline
        1& 1102&241.60 &-21.93 &$-1.09\pm0.29$&$10.15\pm0.20$\\
        2& 454&245.68 &-25.12 &$-0.25\pm0.17$&$12.10\pm0.23$\\
        \hline
        \hline
        H&102&240.6&-22.4& $-3.40\pm0.42$ & $3.96\pm0.21$ \\
        I&110&246.4&-23.9& $-0.78\pm0.91$ & $3.73\pm0.37$ \\
        \hline
        \hline
    \end{tabular}
    \tablefoot{Number of stars (N) and mean positions (RA, Dec) are provided along with the time of gas expulsion ($t_K$, kinematic age) and half-mass radius of subcluster at the time of gas expulsion ($r_{hm}).$}
    \label{tab:subgroups}
\end{table}

\end{appendix}

\end{document}